\documentclass[a4paper,11pt]{article}

\pdfoutput=1

\usepackage{amsmath}
\usepackage{amsfonts}
\usepackage{amssymb}
\usepackage{cite}
\usepackage{multicol,multirow}
\usepackage{graphicx, physics, subcaption}
\usepackage{graphics, epsfig}
\usepackage{epsf}
\usepackage{epstopdf}
\usepackage[usenames,dvipsnames]{color}

\usepackage{colortbl}
\definecolor{summersky}{cmyk}{0.71,0.33,0,0.14}
\definecolor{flamingo}{cmyk}{0,0.51,0.71,0.14}
\definecolor{rp}{cmyk}{0.2, 1, 0.6, 0}
\definecolor{pacificblue}{cmyk}{0.95,0.3,0, 0.19}
\definecolor{gray60}{cmyk}{0.4,0.4,0,0.8}

\numberwithin{equation}{section}

\usepackage[colorlinks,bookmarks=false, hypertexnames=true]{hyperref}
\hypersetup{
	breaklinks=true,
	pdfstartview={FitH},    
	colorlinks=true,       
	linkcolor=blue,          
	citecolor=red,        
	filecolor=magenta,      
	urlcolor=red,           
	anchorcolor=green,      
	linktocpage=true
}

\newcommand{\nc}{\newcommand}

\nc{\ba}{\begin{eqnarray}}
\nc{\ea}{\end{eqnarray}}

\nc{\calR}{{\cal{R}}}
\nc{\calP}{{\cal{P}}}
\nc{\cN}{ {\cal{N}} }

\def\bfk{{\bf k}}
\def\bfq{{\bf q}}
\def\bfm{{\bf m}}

\newcommand{\bfp}{{\bf{p}}}

\begin{document}

\def\thefootnote{\fnsymbol{footnote}}

\begin{center}

\large{\bf  
Hamiltonians to all Orders in Perturbation Theory  and\\ Higher Loop Corrections in  Single Field Inflation with PBHs Formation 
}
\\[0.5cm]

{
Hassan Firouzjahi\footnote{firouz@ipm.ir}, Bahar Nikbakht\footnote{bahar.nikbakht@ipm.ir}
}
\\[0.5cm]
 
 {\small \textit{School of Astronomy, Institute for Research in Fundamental Sciences (IPM) \\ P.~O.~Box 19395-5746, Tehran, Iran
}}\\

\end{center}

\vspace{.3cm}
\hrule
\begin{abstract}

We calculate the action and the interaction Hamiltonians to all orders in perturbation theory in the model of single field inflation with a transient ultra slow-roll phase.  Employing the formalism of  EFT of inflation, we obtain a compact non-perturbative expression for the interaction Hamiltonian in terms of the Goldstone  field $\pi$ in the decoupling limit.  In addition, we also present  a non-linear relation between $\pi$ and the curvature perturbations to all orders in perturbation theory.  These are powerful results which enable us to calculate the cosmological correlators and loop corrections to any order in perturbation theory. As a non-trivial example, we calculate the $L$-loop corrections on long CMB scale perturbations in the USR models which are used for PBHs formation. We show that the loop corrections scale like 
$(\Delta N {\cal P}_e L) ^L$ in which ${\cal P}_e$ is the peak of the power spectrum and $\Delta N$ is the duration of the USR phase. This indicates  that 
the loop corrections grow quickly out of perturbative control for large values of $L$. In the conventional USR setup for PBHs formation with $\Delta N \simeq 2.5$, this happens at  $L=4$.

\end{abstract}

\vspace{0.3cm}
\hrule

\newpage

\section{Introduction}
\label{sec:intro}

To study the predictions of models of inflation in cosmological observables
such as power spectrum and bispectrum,  one requires to calculate 
the action for the curvature perturbations and the corresponding 
interaction Hamiltonians to the desired order. To calculate the power spectrum,  the action to second order in cosmological perturbations is required.   
However, to calculate the bispectrum and higher order correlators, one needs to calculate the cubic and the higher  order actions. Due to non-linearity of GR,
this is a non-trivial task. The first complete analysis to calculate the cubic action in perturbation theory was performed by Maldacena \cite{Maldacena:2002vr} who studied the three-point correlation function in the models of single field slow-roll (SR) inflation.   Going to quartic and higher orders while keeping all perturbations  proved to be challenging, for earlier works concerning quartic action see \cite{Jarnhus:2007ia, Arroja:2008ga}.  

One technical complexity associated with calculating higher order actions and interaction Hamiltonians is the contributions of the non-dynamical fields (i.e. constrained fields) such as the lapse and shift functions in ADM formalism. One has to solve them algebraically to the desired order and plug them back in the action of the curvature perturbations. However, since these fields are gravitationally interacting they may be ignored to leading orders in slow-roll  approximation. The formalism of effective field theory (EFT) of inflation \cite{Cheung:2007st, Cheung:2007sv}  is particularly useful in the decoupling limit where one can neglect these perturbations. Intuitively speaking, in the EFT approach one can consider the contributions of the matter field fluctuations which have the dominant effects in cosmological correlators. The EFT formalism is particularly useful in setup like ultra slow-roll (USR) inflation when the leading interactions are guaranteed to come from the matter field perturbations.

Recently, there have been intense debates on the origin of the large loop corrections in models of single field USR inflation to generate primordial black holes (PBHs) as the candidate for dark matter \cite{Kristiano:2022maq, Kristiano:2023scm, Riotto:2023hoz, Riotto:2023gpm, Choudhury:2023vuj,  Choudhury:2023jlt,  Choudhury:2023rks, Choudhury:2023hvf, 
Choudhury:2024one, Choudhury:2024aji, Firouzjahi:2023aum, Motohashi:2023syh, Firouzjahi:2023ahg, Tasinato:2023ukp, Franciolini:2023agm, Firouzjahi:2023btw, Maity:2023qzw, Cheng:2023ikq, Fumagalli:2023loc, Nassiri-Rad:2023asg, Meng:2022ixx, Cheng:2021lif, Fumagalli:2023hpa,  Tada:2023rgp,  Firouzjahi:2023bkt, Iacconi:2023slv, Davies:2023hhn, Iacconi:2023ggt, Kawaguchi:2024lsw, Braglia:2024zsl, Firouzjahi:2024psd, Caravano:2024moy, Caravano:2024tlp, Saburov:2024und, Ballesteros:2024zdp, Firouzjahi:2024sce,Sheikhahmadi:2024peu, Frolovsky:2025qre,  Kristiano:2024vst, Kristiano:2024ngc}. In the simplest setup, these models involve an intermediate phase of USR sandwiched  between two long  SR phases. The curvature perturbations associated to the modes which leave the horizon during the USR phase experience a growth. It was argued by Kristiano and Yokoyama  \cite{Kristiano:2022maq} that these small scale perturbations can back-react on long CMB scale perturbations via quantum loop corrections. It was argued that the one-loop corrections by these USR modes can significantly affect the long mode power spectrum, destroying the perturbativity of the system. This question was revisited in recent literature with conflicting conclusions \cite{Fumagalli:2023hpa, Tada:2023rgp, Firouzjahi:2023bkt, Inomata:2024lud, Kawaguchi:2024rsv, Fumagalli:2024jzz, Kristiano:2024vst, Kristiano:2024ngc}. 
To study the full one-loop corrections, one needs both the cubic and quartic interaction Hamiltonians. As discussed above, the cubic action is already at hands from Maldacena's analysis while there was no concrete results for the quartic action, specially in a setup containing a USR phase. Fortunately, this question can be addressed nicely within the formalism of EFT of inflation when the cubic and quartic actions of curvature perturbations can be calculated in reasonable ease in decoupling limit. Employing the EFT formalism, the full one-loop corrections on long CMB scale perturbations were performed in \cite{Firouzjahi:2023aum}. The analysis of \cite{Firouzjahi:2023aum} confirmed the conclusion in \cite{Kristiano:2022maq} 
that the one-loop corrections from USR modes can be large so the system may not be under perturbative control. This conclusion was further clarified by including the effects of the sharpness of the transition from the USR phase to the final attractor phase \cite{Firouzjahi:2023aum}.  

Motivated by the above discussions, it is natural to calculate the two and higher orders loop corrections. Needless to say, this will be more complicated than the case of one-loop corrections. The reasons are that there are many more Feynman diagrams in higher loop calculations and also because one has to calculate the higher order interaction Hamiltonians beyond the quartic order. 
As a first attempt, the two-loop corrections associated to the ``double scoop" Feynman diagram involving only the quartic Hamiltonian were calculated 
in \cite{Firouzjahi:2024sce}.  The conclusion of  \cite{Firouzjahi:2024sce} was that the two-loop corrections associated to this Feynman diagram scale like the square of the one-loop correction. This is physically reasonable and, once more,  highlights  the danger of losing perturbative control. To calculate the full 
two-loop corrections, one has to consider  a total 11 one-particle irreducible  Feynman diagrams. This also requires calculating the  quintic and sextic Hamiltonians. 

In this work, we study the action to all orders in perturbation theory within the formalism of EFT of inflation. We are able to obtain a non-perturbative expression for the  interaction Hamiltonian, valid to all orders in perturbation theory. Equipped with this non-perturbative Hamiltonian, we calculate the
$L$-loop corrections in power spectrum associated to a subset of Feynman diagrams involving a single vertex with $L$ loops attached. We obtain the leading loop corrections with interesting implications. A summary of our main results here has been  reported recently in a  short companion paper \cite{Firouzjahi:2025gja}. 

The rest of the paper is organized as follows. In section \ref{SETUP} we briefly review the USR setup. In section \ref{action} we present our results for the interaction Hamiltonian in EFT formalism. In section \ref{loop-correction} we employ our results of  section \ref{action} for the in-in formalism of  loop corrections in the conventional setup used for PBHs formation. 
In sections \ref{bulk} and \ref{boundary}  we calculate in great details the contributions from the bulk of USR as well as  from the localized sources respectively while the total loop correction is given in section \ref{total}.  The summary and discussions are presented in  section \ref{summary} followed by some Appendices containing technical details.

\section{The USR Setup} 
\label{SETUP}

Here we briefly review the USR setup. The model of single field 
USR inflation \cite{Kinney:2005vj} became popular as a counter example violating the celebrated Maldacena non-Gaussianity consistency condition in single field models \cite{Namjoo:2012aa}. In the USR setup the system does not reach to its attractor limit so the cosmological observables  depend on their initial conditions in the phase space. More specifically, during the USR phase the first SR parameter $\epsilon \equiv-\dot H/H^2$ falls off exponentially in which $H$ is the Hubble expansion rate during inflation and a dot denotes the derivative with respect to the cosmic time. Consequently,  the comoving curvature perturbation $\calR$ grows on the superhorizon scales  with the scale factor like $a(t)^3$.  The growth of the curvature perturbations on superhorizon scales is the prime reason for the violation of the single field  non-Gaussianity consistency condition. In the simplest setup studied in 
\cite{Namjoo:2012aa}, the amplitude of non-Gaussianity is obtained to be 
$f_{NL}=\frac{5}{2}$.  For further reviews on various aspects of USR models see \cite{Kinney:2005vj, Namjoo:2012aa, Martin:2012pe, Chen:2013aj, Morse:2018kda, Lin:2019fcz, Dimopoulos:2017ged, Chen:2013eea, Akhshik:2015nfa, Akhshik:2015rwa, Mooij:2015yka, Bravo:2017wyw, Finelli:2017fml, Passaglia:2018ixg, Pi:2022ysn, Ozsoy:2021pws,  Firouzjahi:2023xke, Namjoo:2023rhq, Namjoo:2024ufv, Cai:2018dkf}.

Of course, in order to prevent the curvature perturbation from growing beyond the perturbative regime, one has to terminate the USR phase so the system can follow an attractor SR phase afterwards during which the power spectrum is frozen on superhorizon scales. The mechanism of gluing a USR phase to the final SR phase can be non-trivial which can affect the outcome of the cosmological observables. In the original toy model studied in \cite{Namjoo:2012aa}, it is assumed that the transition from the USR phase to the final attractor phase is extremely sharp, so the mode function is frozen immediately after the USR phase. This simplification assumption was further investigated  in 
\cite{Cai:2018dkf} in which a relaxation period for the evolution of the mode function after the USR phase is considered. It is shown that the sharpness of the transition  can affect the power spectrum and the bispectrum. 
In particular, if the transition is mild so the mode function keeps evolving after the USR phase, then much of the non-Gaussianity is washed out. The maximum allowed value of $f_{NL}$ happens for the case of extreme sharp transition considered in \cite{Namjoo:2012aa} with $f_{NL}=\frac{5}{2}$.
The effects of the sharpness of the transition on the amplitude of loop corrections were also highlighted  in \cite{Firouzjahi:2023aum} where it is shown that the one-loop correction grows (almost) linearly with the sharpness parameter. Consequently,  the danger of large loop corrections  is more severe for the setups where the  transition is sharp (shorter relaxation periods)  \cite{Firouzjahi:2023aum}.  

For more quantitative discussions,  consider the inflaton field $\phi$ in the FLRW metric, 
\ba
ds^2 = -dt^2 + a(t)^2 d{\bf x}^2 \, .
\ea
During the USR phase the inflaton potential is exactly flat, $V(\phi)= V_0$, and  the background equations are, 
\ba
\ddot \phi(t) + 3 H \dot \phi(t)=0\, , \quad \quad 3 M_P^2 H^2 \simeq V_0, 
\ea
in which in the second equation we have neglected the kinetic energy of the inflaton field compared to $V_0$. Here, 
$M_P$ is the reduced Planck mass while  $H$ is the Hubble expansion rate which is very nearly constant during inflation.  To have a consistent inflationary setup, we assume that the USR phase is terminated at a moment when $\phi=\phi_e$.   
Since the potential is flat, then  $\dot \phi \propto a^{-3}$ and    
$\epsilon = -{\dot H}/{H^2} $ decays like $a^{-6}$. On the other hand, the other  SR parameter $\eta \equiv {\dot \epsilon}/{H \epsilon}$ attains a large negative value $\eta=-6$, which is the hallmark of the USR phase.

The growth of the curvature perturbations is associated to the non-trivial behaviour of the would-be decaying mode. More specifically, the would-be decaying mode  becomes the growing mode during the USR phase along with the other mode which is a constant similar to other single field models. To see this,  consider the evolution of the curvature perturbation $\calR$ in Fourier 
space \cite{Bassett:2005xm, Abolhasani:2019cqw}, 
\ba
\big( a^2 \epsilon \calR')' + k^2 a^2 \epsilon \calR=0 \, ,
\ea
in which a prime indicates the derivative with respect to the conformal 
time $\tau$ which is  related to the cosmic time via $d\tau= dt/a(t)$.
On the superhorizon scales, this equation can be solved easily, yielding,
\ba
\label{R-sol}
\calR= C_1+ C_2 \int \frac{d \tau}{a^2 \epsilon} \, ,
\ea
in which $C_1$ and $C_2$ are two constants of integration, representing two independent modes of solution.  The key point is that during the USR phase  $\epsilon \propto a^{-6}$ so the mode involving $C_2$ grows like $a(t)^3$ as discussed before. This is in contrast to conventional SR model in which $\epsilon$ is nearly constant so this mode decays quickly and the constant mode $C_1$ represents the dominant solution. 

Depending on model building, the USR phase can be either an extended or 
an intermediate one. In the former, as originally studied in \cite{Namjoo:2012aa}, inflation has only two phases. The USR phase is the earlier phase 
of inflation where the CMB scale perturbations leave the horizon followed by a SR phase.  As the long modes leave the horizon during the USR phase, a large non-Gaussianity is generated on these scales which can be observed in CMB maps.  However, more recently, it becomes popular to consider a setup in which the USR phase is an intermediate one, sandwiched between two SR phases. These setups were employed to generate PBHs as a candidate for dark matter
\cite{Ivanov:1994pa, Garcia-Bellido:2017mdw, Germani:2017bcs, Biagetti:2018pjj}, for a review on  PBHs from USR setup see \cite{Khlopov:2008qy, Ozsoy:2023ryl, Byrnes:2021jka, Escriva:2022duf, Pi:2024jwt}. In this latter setup, the USR phase is assumed to be extended in the  time 
interval $\tau_s < \tau <\tau_e$ so $\epsilon$ at the end of USR is related to its value at the start of USR via  $\epsilon_e = \epsilon_i \big( \frac{\tau_e}{\tau_s} \big)^6 $. As we shall see, it is convenient to work with the number of e-folds $N$ as the clock in which  $d N= H dt$. In terms of $N$ we have 
 $\epsilon_e = \epsilon_i e^{-6 \Delta N}  $ in which $\Delta N \equiv N(\tau_e) - N(\tau_s)$  represents the duration of the USR phase. 
 
The  power spectrum at the end of USR phase is given by, 
\ba
\label{powr-P}
P_{\calR}(\bfk, \tau_e) =  | \calR(\tau_e) |^2 = 
\frac{ H^2}{4 \epsilon_e  k^3 M_P^2} \, .
\ea
To generate PBHs with the desired mass scales, it is usually  assumed that 
$P_{\calR}(\bfk, \tau_e) $ has increased by about 7 orders of magnitude compared to its CMB scale amplitude, i.e. $\epsilon_e/\epsilon_i \simeq 10^{-7}$. Solving this for $\Delta N$, we obtain  $ \Delta N \simeq 2.5$ e-folds while its exact value is model dependent. 

Gluing the USR phase to the final attractor phase plays important roles. In our analytical investigation, we consider the simple picture in which the transition from the USR phase to the final SR phase happens instantaneously at the time 
$\tau=\tau_e$. However, as discussed above, the mode function may keep evolving after $\tau > \tau_e$ before it assumes its final fixed values. To capture this property, the sharpness parameter (or the relaxation) parameter 
$h$ was introduced in  \cite{Cai:2018dkf} via, 
\ba
\label{h-def}
h
=-6 \sqrt{\frac{\epsilon_V}{\epsilon_e}} \, .
\ea
Here, $\epsilon_V$, constructed from the  first derivative of the potential, represents the value of $\epsilon$ when the system reaches its attractor phase. Without loss of generality, we assumed that $\dot \phi<0$ so $h<0$. For a  sharp transition $|h| >1$ while for a mild transition $h$ is comparable to  slow-roll parameters. In the former case the mode function is frozen quickly after the USR phase while in the latter situation the mode function keeps evolving for an extend period after the USR phase.  As a benchmark example, for the instant sharp transition considered  in \cite{Kristiano:2022maq}, we have  $h=-6$.  This corresponds  to the situation in which  $\epsilon$ during the final SR stage  is frozen to its value at the end of USR with 
 $\epsilon_V =\epsilon_e$. In our analysis based on EFT approach in which we neglect the slow-roll corrections, we assume that the transition is sharp enough with $|h| >1$ so the relaxation period immediately after the USR phase can be neglected. 

As shown in \cite{Cai:2018dkf}, the evolution of  the second slow-roll parameter $\eta $ near the 
transition point is  approximated via,  
\ba
\label{eta-jump1}
\eta = -6 - h \theta(\tau -\tau_e) \quad \quad  \tau_e^- < \tau < \tau_e^+ \, ,
\ea
which gives,     
\ba
\label{eta-jump}
\frac{d \eta}{d \tau} = - h \delta (\tau -\tau_e)  \, ,  \quad \quad  \tau_e^- < \tau < \tau_e^+ \, .
\ea 
From the above expressions we see that $\eta$ experiences a jump at the 
time $\tau=\tau_e$ which is highlighted by $\delta (\tau -\tau_e)$ in $\eta'$.
This jump in $\eta$ will have important effects in the interaction Hamiltonian and the loop corrections as we shall see in the following analysis.

In our analytical investigation,  we  employ the simple picture considered above in which the transition from the USR phase to the SR phase happens instantaneously at $\tau=\tau_e$. However, in a realistic situation this transition does not take place instantaneously and will take some time. In addition, we consider the setup in which the relaxation period associated with the evolution of the mode functions is short, corresponding to $|h|\gg 1$. These two are the main simplification assumptions which are considered in this work. Studying the more realistic picture, in which the transition from the USR phase to the SR phase is smooth and the relaxation time for the evolution of the mode function is long enough, requires numerical investigation. However, this also makes the theoretical analysis, such as the in-in analysis, intractable. Having said this, one should keep in mind the difference  between the instantaneity of the transition and the sharpness of the transition. Specifically, the instantaneity of the transition assumes that the gluing of the USR to the SR happens at a point, $\tau=\tau_e$. Even in this case of instant transition, the evolution of  mode function 
can be mild in the sense that the mode function can keep evolving for a long period before reaching its attractor value corresponding to 
 $|h|\ll 1$.

\section{ Action and Hamiltonian to all Orders in EFT} \label{action}

In this section we present our results for the action of cosmological perturbations and the interaction Hamiltonian to all orders in perturbation theory. To perform the analysis, we employ the EFT formalism of inflation \cite{Cheung:2007st, Cheung:2007sv} which is proved very successful in USR setup. While our main interest is in the USR model with $\eta=-6$, but the analysis here is valid for  a general constant-roll setup in which $\eta$ is constant. The EFT approach in USR setup was first employed  in \cite{Akhshik:2015nfa} who studied the bispectrum and the shape of the non-Gaussianity in a general non-attractor model including the USR setup. The EFT approach was recently employed in \cite{Firouzjahi:2023aum, Firouzjahi:2024sce}  as well to calculate the quartic Hamiltonian for the  loop corrections in the setup considered in \cite{Kristiano:2022maq}, see also \cite{Fumagalli:2024jzz} for the quartic Hamiltonian.  

In a general single field inflation model, the cosmological perturbations are encoded in both metric perturbations $\delta g_{\mu \nu}$ and the inflaton field perturbation $\delta \phi$. In the absence of perturbations, the background fields are homogenous but time-dependent. The spacetime has the FLRW metric while the inflaton field has the time-dependent profile $\phi=\phi(t)$. As a result,  the full four-dimensional diffeomorphism invariance is broken to a subset of three-dimensional diffeomorphism invariance. This is the key step in constructing the action of cosmological perturbations in EFT formalism.  More specifically, to construct the action of cosmological perturbations, we first go to the unitary (comoving) gauge where the inflaton perturbation is turned off while all perturbations are encoded in the metric sector. In this gauge, one writes down all operators which are consistent with the remaining three-dimensional diffeomorphism invariance. After writing the full action, we move away from the comoving gauge by 
turning on the inflaton perturbation  via $\pi(x^\mu)$, the Goldstone boson associated to the breaking of the time diffeomorphism invariance. 

Up to this stage, everything was general and EFT formalism dos not provide particular simplification in the analysis compared to the conventional approach. However, a significant simplification arises when one goes to the decoupling limit where the lapse and shift functions in ADM formalism are trivial, i.e. $N=1$ for the lapse function and 
$N^i=0$ for the shift function. In general, the lapse and shift functions are non-dynamical variables with algebraic equations which should be solved in favour of the curvature perturbation. This process is messy which makes the analysis 
complicated  when one goes to higher orders in perturbation theory \cite{Maldacena:2002vr}. However,  these non-dynamical fields are only gravitationally interacting, and their amplitudes are higher orders in powers  of $\epsilon$, which are negligible. Therefore, in setups like USR  or  $P(X, \phi)$ model where the dominant interactions come from the matter sector, one may safely go to the decoupling limit and neglect the contributions of non-dynamical fields. This is the advantage of the EFT formalism which simplifies the analysis immensely as we shall show below.

As elaborated in \cite{Cheung:2007st, Cheung:2007sv}, the full action is 
$S=S_{\mathrm{matter}}+ S_{\mathrm{EH}}$ in which $S_{\mathrm{EH}}$ is the usual Einstein-Hilbert action while $ S_{\mathrm{matter}}$ is the action associated to the matter sector. In the decoupling limit, the metric is unperturbed and all perturbations are encoded in the inflaton sector.  Assuming a canonically normalized scalar field with the sound speed $c_s=1$, the matter action in a given FLRW background is given by, 
\begin{equation}
\label{eq:Smatter Full}
\begin{aligned}
S_{\text {matter }}= \int d^4 x \sqrt{-g} & {\left\{-M_P^2 \dot{H}(t+\pi)\left[\frac{1}{N^2}\left(1+\dot{\pi}-N^i \partial_i \pi\right)^2-\partial^i \pi \partial_i \pi\right]\right.} \\
&~~~~~  \left.-M_P^2\left[3 H^2(t+\pi)+\dot{H}(t+\pi)\right]\right\},
\end{aligned}
\end{equation}
where at this stage we have kept the lapse and the shift functions $N$ and $N^i$ for comparison. The error in working in the decoupling limit, 
where the the metric perturbations $N$ and  $N^i$ can be neglected, is 
at the order  ${\cal O}(\epsilon^2)$. We have checked explicitly that the decoupling limit does hold  in the USR setup to seven orders in perturbation theory. Having said this, from the structure of the equations for 
$N$ and $N^i$, one expects that the decoupling limit to hold to all orders in perturbation theory   \cite{Green:2024hbw}. 

Now, going to the decoupling limit and setting $N = 1$ and $N^i = 0$ as explained above,  Eq.~\eqref{eq:Smatter Full} simplifies as follows,
\begin{equation}
\label{eq:Smatter}
S_{\text {matter }}= - M_P^2 \int d^4 x \,  a^3(t) \left[3 H^2(t+\pi) + 2 \dot{H}(t + \pi) + 2 \dot{H}(t + \pi) \dot{\pi} + \dot{H}(t + \pi) \left( \dot{\pi}^2 - \partial_i \pi \partial^i \pi \right) \right] \, .
\end{equation}
In this limit, all interactions are encoded in background functions  $H(t+\pi)$ and $\dot H(t+\pi)$.

To calculate the interaction Hamiltonian at the $n$-th order in perturbations, we need the $n$-th order action. Expanding $H(t+\pi)$ and $\dot H(t+\pi)$ in Eq. (\ref{eq:Smatter}) to that order, we obtain, 
\begin{equation}
\begin{aligned}
\label{eq:Sn}
S_{\pi^n} = - M_P^2 \int d^4x\,  a^3(t)\,  \left\{ 3 \left[ H^2(t+\pi) \right]_n + 2 \left[ \dot{H}(t + \pi) \right]_n + 2 \left[ \dot{H}(t+\pi) \right]_{n-1} \dot{\pi} \right. 
\\ 
+\left.  \left[ \dot{H}(t+\pi) \right]_{n-2 }  \left( \dot{\pi}^2 - \partial_i \pi \partial^i \pi \right) \right\} \, ,
\end{aligned}
\end{equation}
where $\left[ ... \right]_n$ represents the terms of order $n$ in the 
expansion. Here $S_{\pi^n}$ represents the $n$-th order of the matter action 
which is simply denoted by $S$.  

It can be shown that,
\ba
\label{H-app1}
\left[ H^2(t+\pi) \right]_n  &=& \frac{2}{n !} H(t) H^{(n)}(t) \pi^n + \mathcal{O}(\epsilon^2)\, ,\\
\label{H-app2}
\left[ \dot{H}(t + \pi) \right]_n &=& \frac{1}{n!} H^{(n+1)} (t) \pi^n,
\ea
where $H^{(n)}(t)$ is the $n$-th derivative of the Hubble parameter with respect to  the cosmic time. In particular, $H^{(1)} = \dot H= -\epsilon H^2$, while
\begin{eqnarray}
{H^{(2)}}&=&-\dot{\epsilon}H^2-2\epsilon \dot{H}H \, , \nonumber \\
&=&-\eta \epsilon H^3 + 2\epsilon ^2 H^3
= -\eta \epsilon H^3 + O(\epsilon ^2),
\end{eqnarray}
and for higher orders of $H^{(n)}$, similarly we have  \cite{Firouzjahi:2023aum}, 
\begin{eqnarray}
\label{H3-eq}
{H^{(3)}} &=& -\eta ^2 \epsilon H^4 - \epsilon  \frac{d \eta}{d t} H^3 + O(\epsilon ^2) \, ,\\
\label{H4-eq}
{H^{(4)}} &=& -\eta ^3 \epsilon H^5 - 3 \epsilon  \eta  \frac{d \eta}{d t} H^4
- \epsilon \frac{d^2 \eta}{d t^2} H^3
+O(\epsilon ^2) \, , \\
\label{H5-eq}
{H^{(5)}} &=& -\eta ^4 \epsilon H^6 - 6 \epsilon  \eta^2  \frac{d \eta}{d t} H^5
- 3 \epsilon  \left( \frac{d \eta}{d t}\right)^2 H^4 - 4 \epsilon  \eta  \frac{d^2 \eta}{d t^2} H^4
-   \epsilon    \frac{d^3 \eta}{d t^3} H^3
+O(\epsilon ^2) \, .
\end{eqnarray}
It is crucial to note that $H^{(n)} \sim -\epsilon H^{n+1} \eta^{n-1}$ plus terms of higher orders in $\epsilon$ which can be neglected. Also, note that for $n \ge 3$,  $H^{(n)}$
involves  $\dot \eta, \ddot \eta, \dot \eta^2 $ etc. Since $\eta$ has a jump at $t=t_e$, these terms involve the Dirac delta functions $\delta (t -t_e)$ and its derivatives. This will bring major complexities at the point of end of USR as we will elaborate in next section when studying the loop corrections.  

Now plugging Eqs. (\ref{H-app1}) and (\ref{H-app2}) in  
Eq. (\ref{eq:Sn}),  the expansion of the action takes the following form, 
\begin{equation}
\label{eq:Sn Full}
\begin{aligned}
S_{\pi^n} =& -M_P^2 \int d^4x\,  a^3(t) \left[ \frac{6}{n!} H(t) H^{(n)}(t) \pi^n + \frac{2}{n!} H^{(n+1)}(t) \pi^n + \right.\\
&\left. \frac{2}{(n-1)!} H^{(n)}(t) \pi^{n-1} \dot{\pi} + \frac{1}{(n-2)!} H^{(n-1)}(t) \pi^{n-2}  \left( \dot{\pi}^2 - \partial_i \pi \partial^i \pi \right) \right]+ {\cal O}(\epsilon^2) \, .
\end{aligned}
\end{equation}

On the other hand, note that 
\begin{equation}
\frac{2}{n!} \frac{d}{dt} \left( a^3(t) H^{(n)}(t) \pi^n \right) = a^3(t) \left[\frac{6}{n!} H(t) H^{(n)}(t)\pi^n + \frac{2}{n!} H^{(n+1)}(t) \pi^n + \frac{2}{(n-1)!} H^{(n)} \pi^{n-1} \dot{\pi} \right].
\end{equation}
Thus Eq.~\eqref{eq:Sn Full} can be simplified as follows,
\begin{equation}
\label{eq:Sn final0}
S_{\pi^n} = -M_P^2 \int  d^4x\,  a(t)^3 \left[ \frac{1}{(n-2)!} H^{(n-1)}(t) \pi^{n-2}  \left( \dot{\pi}^2 - \partial_i \pi \partial^i \pi \right) \right] + \frac{2}{n!} \frac{d}{dt} \left(a^3(t) H^{(n)}(t) \pi^n \right)  \, .
\end{equation}
 An important point to observe is that  the total time derivative term in Eq.~\eqref{eq:Sn final0}  is merely a function of $\pi$ so we can discard it safely at the boundaries $t\rightarrow 0$ and $t\rightarrow t_0$ in which $t_0$ is the time of end of inflation. It is important that the total derivative term in  Eq.~\eqref{eq:Sn final0}  does not depend on $\dot \pi$. As shown in \cite{Braglia:2024zsl}, total derivative terms 
involving only functions of $\pi$ and not $\dot \pi$ are harmless as they can be absorbed via appropriate canonical transformations in the phase space spanned by $(\pi, \dot \pi)$.

Thus, neglecting the total time derivative term,  the $n$-th order action simplifies to,
\begin{equation}
\label{eq:Sn final}
S_{\pi^n} = -M_P^2 \int  d^4x\,  a(t)^3 \left[ \frac{1}{(n-2)!} H^{(n-1)}(t) \pi^{n-2}  \left( \dot{\pi}^2 - \partial_i \pi \partial^i \pi \right) \right]  + {\cal O}(\epsilon^2) \quad \quad (n\ge 2) \, .
\end{equation}
One can check that the action \eqref{eq:Sn final} matches exactly 
with the result obtained in \cite{Firouzjahi:2023aum} up to fourth order which was required in \cite{Firouzjahi:2023aum} for one-loop calculations. 

Having obtained the $n$-th order of the action, we can combine them to form the total action which reads, 
\ba
S &=& \sum_{n=2} S_{\pi^n}= -M_P^2 \int  d^4x\,  a(t)^3 
\sum_{n=2} \left[ \frac{1}{(n-2)!} H^{(n-1)}(t) \pi^{n-2}\right]   \left( \dot{\pi}^2 - \partial_i \pi \partial^i \pi \right) +  {\cal O}(\epsilon^2) \nonumber\\
&=& -M_P^2 \int  d^4x\,  a(t)^3 \dot H(t+ \pi)  \left( \dot{\pi}^2 - \partial_i \pi \partial^i \pi \right) + {\cal O}(\epsilon^2) \,  . 
\ea
Curiously, we see that the action is simply the last term in 
Eq. (\ref{eq:Smatter}) while the first three terms in Eq. (\ref{eq:Smatter}) assemble themselves into a total time derivative term plus a correction of order ${\cal O}(\epsilon^2)$ which can be neglected to leading order in slow-roll approximation.   

Finally, using the definition of $\epsilon$, we can replace $\dot H(t+\pi)$ in favour of $\epsilon(t+ \pi)$, obtaining the following form of the action,
\ba
\label{action-pi}
S= M_P^2 H^2 \int  d^4x\,  a(t)^3 \epsilon(t+ \pi)  \left( \dot{\pi}^2 - \partial_i \pi \partial^i \pi \right) + {\cal O}(\epsilon^2) \, .
\ea
The above action is valid to all orders in perturbation theory in the decoupling limit for any single field model with $c_s=1$.   As a special case, now consider the bulk of USR (or a general constant-roll) with no boundary where  $\eta=\frac{\dot \epsilon}{H \epsilon}$ is constant without any discontinuity in $\eta$.   In this case $\epsilon(t)\propto e^{\eta H t}$ so $\epsilon(t+\pi)= \epsilon(t) e^{\eta H \pi}$. Plugging this into action (\ref{action-pi}),  we obtain the following compact form of the action,
\ba
\label{action-pi2}
S= M_P^2 H^2 \int  d^4x\,  a(t)^3 \epsilon(t)\,  e^{\eta H \pi}  \left( \dot{\pi}^2 - \partial_i \pi \partial^i \pi \right) + {\cal O}(\epsilon^2) \, ,  \quad \quad 
(\eta= \mathrm{constant}) \, .
\ea
The fact that such a simple form of the non-perturbative action can be obtained  is a demonstration of the strength of EFT formalism in the decoupling limit.  

\subsection{Non-Perturbative Interaction Hamiltonian}
\label{Hamiltonian}

To calculate the interaction Hamiltonian care must be taken as 
we must first define the conjugate momentum  associated to  $\pi$.
From the  action  (\ref{action-pi}), the conjugate momentum is given by, 
\begin{equation}
\label{eq:Momentum Density}
\Pi \equiv  \frac{\partial \mathcal{L}}{\partial \dot{\pi} } = 
2M_P^2 H^2 a^3 \epsilon(t+ \pi)  \dot \pi  + {\cal O}(\epsilon^2)\, .
\end{equation}
Having calculated the conjugate momentum $\Pi$, and discarding the 
${\cal O}(\epsilon^2)$ corrections,  the Hamiltonian density  is given by, 
\begin{equation}
\label{eq:Hamiltonian}
\mathcal{H} = \Pi \dot{\pi} - \mathcal{L} = \frac{\Pi^2}{4 a^3 M_P^2 H^2 \epsilon(t+\pi)} + M_P^2 H^2 a^3 \epsilon(t+ \pi) \partial_i \pi \partial^i \pi 
+ {\cal O}(\epsilon^2) \, .
\end{equation}

The above expression for the Hamiltonian is exact and it contains the 
Hamiltonian of the free theory as well as the interacting Hamiltonian. To construct the  interaction Hamiltonian, we should identify the Hamiltonian of the free theory $\mathcal{H}_0$ and subtract it from the total Hamiltonian. 
The Hamiltonian of the free theory is,
\ba
\mathcal{H}_0 = \frac{\Pi^2}{4 a^3 M_P^2 H^2 \epsilon(t)}
+ M_P^2 H^2 a^3 \epsilon(t) \partial_i \pi \partial^i \pi \,  + {\cal O}(\epsilon^2).
\ea 
Subtracting it from the total Hamiltonian, the interaction Hamiltonian is obtained to be 
\ba
\mathcal{H}_I = \frac{\Pi_I^2}{ 4 M_P^2 H^2 a^3} \Big(\frac{1}{\epsilon(t+ \pi_I)} 
-\frac{1}{\epsilon(t)}\Big)  + M_P^2 H^2 a^3 \Big(  \epsilon(t+ \pi_I) - \epsilon(t) \Big)  \partial_i \pi_I \partial^i \pi_I + {\cal O}(\epsilon^2) \, ,
\ea 
where the subscript $I$ indicates that all fields are calculated in the interaction picture, i.e. with the mode function given by the free theory.

Now substituting  $\Pi_I$ from Eq. (\ref{eq:Momentum Density}) in terms of 
$\dot \pi_I$ and dropping  the subscript $I$ on the interaction picture fields from now on, we end up with the following  form of the total interaction Hamiltonian,
\ba
\label{HI-eq2}
\mathcal{H}_I = M_P^2 H^2 a(t)^3 \Bigg[\epsilon(t)^2  \Big(\frac{1}{\epsilon(t+ \pi )}   -\frac{1}{\epsilon(t)}\Big) \dot \pi^2 +  \Big(  \epsilon(t+\pi) - \epsilon(t) \Big)  \partial_i \pi \partial^i \pi \Bigg] + {\cal O}(\epsilon^2) \, .
\ea

The above form of the interaction Hamiltonian is general and can be applied to any single field models with $c_s=1$ in the decoupling limit. In particular, it can be used in the setup where a USR phase is glued to the SR phase as in the model of \cite{Kristiano:2022maq} employed for PBHs formation. In the setup of \cite{Kristiano:2022maq}  there is a jump in $\eta$ so the expansion of the function $\epsilon(t+\pi)$ at second and higher orders will induce singularities 
in the form of Dirac delta function and its derivatives. We will come back to this case in the context of loop correction in next sections. 

In the special case of the bulk of USR (or a general constant-roll) with no boundary where   $\epsilon(t+\pi)= \epsilon(t) e^{\eta H \pi}$ with $\eta$ being a constant, from  Eq. (\ref{HI-eq2}) we obtain the following compact form of the interaction Hamiltonian,
\ba
\label{HI-eq3}
\mathcal{H}_I = M_P^2 H^2 a^3 \epsilon(t) \Big[  \Big(e^{-\eta H \pi}-1\Big) \dot \pi^2 +  \Big( e^{\eta H \pi}-  1 \Big)  \partial_i \pi \partial^i \pi \Big] + {\cal O}(\epsilon^2)  \quad \quad 
(\eta= \mathrm{constant}) \, .
\ea
Note that the above Hamiltonian can be obtained from the bulk action (\ref{action-pi2}) as well. 

Eq.  (\ref{HI-eq3}) is an interesting  result, providing a non-perturbative form of the interaction Hamiltonian in terms of the Goldstone field $\pi(x^\mu)$. One can check that the  formula (\ref{HI-eq3}) matches exactly with the cubic and quartic Hamiltonians  of \cite{Firouzjahi:2023aum}  in the bulk  when $\dot \eta=0$.  

\subsection{Non-linear Relation Between $\pi$ and $\calR$}
\label{pi-R}

Of course, we are interested in correlation functions of the curvature perturbation $\calR$ so to complete our job we need a dictionary between $\pi$ field and $\calR$. 
To find the relation between $\pi(x^\mu)$ and $\calR$, we consider the comoving and flat gauges as follows \cite{Maldacena:2002vr}.  First, we neglect the spatial dependence of the perturbations which are consistent in superhorizon limit.  
In the comoving gauge, equipped with the time coordinate $t$, the scalar perturbations are given as follows,
\ba
\label{comoving}
\pi(t)=0, \quad \quad h_{ij} = a(t)^2 e^{2 \calR} \delta_{ij} \, ,
\ea 
in which $h_{ij}$ represents the spatial components of the metric. 
On the other hand, the flat gauge is equipped with new time $\tilde t$ in which the perturbations are given as follows,
\ba
\pi(\tilde t) \neq 0 , \quad \quad h_{ij} = a(\tilde t)^2 \delta_{ij} \, .
\ea
The two coordinate systems are related via $ \tilde t= t + T$ in which $T$ is a time diffeomorphism parameter. Relating the metric parts in these two gauges, and discarding 
the spatial derivatives in superhorizon limits  we obtain, 
\ba
\calR = \ln \Big( \frac{a(t+T)}{a(t)} \Big) \, .
\ea 
Using the definition $H =\dot a(t)/a(t)$, the above equation can be expressed as follows,
\ba
\calR = \int_t^{t+T} dt' H(t') \, .
\ea 
Solving this equation perturbatively in powers of $T$, we obtain
\ba
\label{R-eq1}
\calR = \sum_{n=1} \frac{1}{n!}  H^{(n-1)}  \, T^n \, .
\ea

On the other hand, in going from the flat gauge to the coming gauge, we demand that the Goldstone boson to vanish, $\pi(t+T)+ T=0$. This equation has the following solution for $T$ in powers of $\pi$,
\ba
\label{T-eq}
T= \sum_{n=1} \frac{(-1)^n}{n!} \frac{d^{n-1}}{d t^{n-1}}  \pi^n\, .
\ea 

Combining Eqs. (\ref{R-eq1}) and (\ref{T-eq}) and using the binomial and the general Leibniz rule, we finally obtain the relation between $\calR$ and $\pi$ as follows,
\ba
\label{R-eq2}
\calR &=& \sum_{n=1} \frac{(-1)^n}{n!}  \frac{d^{n-1}}{d t^{n-1}} 
\Big( H(t) \pi^n\Big) \\
&=& - H \pi + \frac{1}{2} \frac{d}{dt} \big( H(t) \pi^2 \big)  
- \frac{1}{6} \frac{d^2}{dt^2} \big( H(t) \pi^3 \big) + ... \, .
\ea
This completes our analysis of calculating the interaction Hamiltonian to an arbitrary order in terms of $\calR$ in a USR (or constant-roll) setup where 
$\eta=\mathrm{constant}$.   More specifically, Eq. (\ref{HI-eq2}) provides a non-perturbative expression for the interaction Hamiltonian in terms of $\pi$  while  the $\pi$ field itself can be expressed to all orders  in terms of $\calR$ from  Eq. (\ref{R-eq2}). In practice, to use Eq. (\ref{R-eq2}), we note that the $\pi$ field is in the interaction picture so one can use the free field equation in the superhorizon limit to relate $\dot \pi, \ddot \pi$ etc  in terms of $\pi$. As a result, we end up with an algebraic relation between $\calR$ and $\pi$ which can be solved perturbatively to any desired order. 

We will use the general interaction Hamiltonian  Eq. (\ref{HI-eq2}) in the next section to calculate the loop corrections in the setup of   \cite{Kristiano:2022maq}. 
However, before going to these complicated analysis, here we pause to present a simple application of the above result as
a prelude to the loop analysis.

\subsection{Example: Bispectrum in USR Inflation}
\label{bispectrum}

As a simple application of the above results, here we briefly review the analysis of bispectrum in the extended USR setup which was studied in more details in  \cite{Akhshik:2015nfa} using the EFT formalism. This is a two-step model in which the first phase is a USR which is glued to the SR phase as  proposed in \cite{Namjoo:2012aa}. The transition to the final SR phase is infinitely sharp so the mode function is frozen immediately after the  USR   phase  at $\tau=\tau_e$. This is a simplification which allows one to neglect the evolution of the mode functions after the USR phase. The goal of this simple exercise is to demonstrate that the results for bispectrum obtained within the EFT formalism indeed matches the corresponding results obtained from standard perturbation theory such as from Maldacena's approach or from $\delta N$ formalism.

To calculate the bispectrum we only need the cubic Hamiltonian  ${\bf H}_3$ which from  Eq. (\ref{HI-eq3}) is obtained to be,
\ba
\label{H3}
{\bf H}_3 = - M_P^2 H^3 \eta \epsilon a^2\, \int d^3 x  \Big(  \pi \pi'^2  - \pi (\partial \pi)^2 \Big) \, ,
\ea
where a prime denotes the derivative with respect to the conformal time.

To calculate the three point function of $\calR$ we must take into account the non-linear relation between $\calR$ and $\pi$ to quadratic order 
as well. Using Eq. (\ref{R-eq2}) 
and noting that during the USR phase the free mode function on superhorizon scales is growing with  $\dot \pi_k = -\frac{\eta}{2} H \pi_k$, the non-linear relation between $\calR$ and $\pi$ to quadratic order  is given by,
\ba
\label{R-Pi-2}
\calR = - H \pi -\frac{\eta}{2} H^2 \pi^2  + {\cal O}(\epsilon) \, .
\ea
Correspondingly, the three-point function for the modes $\bfk_1, \bfk_2$ and 
$\bfk_3$ in Fourier space is given by,
\ba
\label{pi-R-bi2}
\Big \langle \calR(\bfk_1)  \calR(\bfk_2) \calR(\bfk_2)  \Big \rangle^\prime_{\mathrm{total}}  &=&  \Big \langle \calR(\bfk_1)  \calR(\bfk_2) \calR(\bfk_2)  \Big \rangle^\prime_{\mathrm{bulk}} + 
\Big \langle \calR(\bfk_1)  \calR(\bfk_2) \calR(\bfk_2)  \Big \rangle^\prime_{\mathrm{boundary}} \, ,
\ea
where it is understood that all quantities are calculated at   the time $\tau_e$ and  $\langle \rangle^\prime$ indicates that we have neglected the common factor $( 2 \pi)^3 \delta^{(3)} (\bfk_1 + \bfk_2 + \bfk_3) $. 

To calculate the first term in the right hand side of Eq. (\ref{pi-R-bi2}) we need the cubic Hamiltonian as given in Eq. (\ref{H3}). Performing the standard in-in analysis in the bulk of USR ($\tau< \tau_e$), one finds \cite{Akhshik:2015nfa}
\ba
\Big \langle \calR(\bfk_1)  \calR(\bfk_2) \calR(\bfk_2)  \Big \rangle^\prime_{\mathrm{bulk}} = \frac{\eta }{2} \Big[ P(\bfk_1)  P(\bfk_2)  +   P(\bfk_1)  P(\bfk_3) +  P(\bfk_2)  P(\bfk_3) \Big] \, .
\ea
On the other hand, using Eq. (\ref{R-Pi-2}), the contribution of the boundary term at $\tau=\tau_e$ is obtained to be \cite{Akhshik:2015nfa},  
\ba
\Big \langle \calR(\bfk_1)  \calR(\bfk_2) \calR(\bfk_2)  \Big \rangle^\prime_{\mathrm{boundary}}= - \eta  \Big[ P(\bfk_1)  P(\bfk_2)  +   P(\bfk_1)  P(\bfk_3) +  P(\bfk_2)  P(\bfk_3) \Big] \, .
\ea

Combining the two contributions from the bulk and the boundary, we obtain 
\ba
\label{bi-total}
\Big \langle \calR(\bfk_1)  \calR(\bfk_2) \calR(\bfk_2)  \Big \rangle^\prime_{\mathrm{total}} =  -\frac{\eta}{2}\Big[\, P(k_1) P(k_2) +  P(k_1) P(k_3) + P(k_2) P(k_3) \, \Big] \, .
\ea
This corresponds to a local-shape non-Gaussianity with the amplitude  $f_{NL}=- \frac{5\eta}{12}$, which for the USR case yields the expected result $f_{NL}= \frac{5}{2}$ obtained from standard methods such as  from Maldacena's approach or from $\delta N$ formalism \cite{Namjoo:2012aa}. 

There are two lessons in this simple example which will be helpful for more complicated cases. First, one should take into account the non-linear relation between $\pi$ and $\calR$ to the appropriate order when one is calculating the correlators. Second, the mode functions in cosmological measurements should be considered at the time of end of inflation. So if the system has not reached its attractor phase immediately after the USR, then its evolution should be taken into account as well. Only in the case of an infinite sharp transition (like in the above example) when the mode function is frozen immediately after the USR phase, one can use the mode function at the end of USR as the final mode function.

\section{Higher Order Loop Corrections in Power Spectrum}
\label{loop-correction}

In this and next sections we employ the results obtained in the previous section to calculate the $L$-th order loop correction in the setup considered 
in \cite{Kristiano:2022maq}.  This is a three-phase setup $\mathrm{SR\rightarrow USR\rightarrow SR}$ in which the curvature perturbation experiences a growth during the USR stage. The USR phase is extended in the interval $\tau_s \leq \tau \leq \tau_e$. As discussed in section \ref{SETUP}, the transition to the final SR phase can be either sharp or mild, depending on the sharpness (relaxation) parameter $h$, for further details see \cite{Firouzjahi:2023aum, Firouzjahi:2024sce}.


To calculate the $L$-th order loop correction, we need the interaction Hamiltonian to higher orders. For example, for one-loop corrections 
($L=1$), we need the interaction Hamiltonian up to quartic order.  The case of one-loop correction, considering only  the cubic interaction, was studied in \cite{Kristiano:2022maq} and in many follow up papers. The complete analysis of one-loop corrections, involving both the cubic and quartic Hamiltonians, were studied in \cite{Firouzjahi:2023aum}. Here we proceed to the case of $L$-loop corrections. 

As one goes beyond one-loop, the number of  one-particle irreducible Feynman diagrams increases rapidly. 
To count the number of independent diagrams, suppose we consider the following scalar interaction,
\ba
V= \sum_n g_n \phi^{n}  \, \quad \quad (n>2)\, ,
\ea
where $g_n$ is the coupling (vertex) while $n$ is the order of the corresponding interaction. For example, $n=3, 4$ for cubic and quartic Hamiltonians respectively.

Now consider a one-particle irreducible diagram with $L$ loops, $V_n$ vertices for the power of interaction $n$,  $P$ internal propagators  and $N$ external lines. Starting with the following topological relations \cite{Weinberg:1995mt},
\ba
L= P- \sum_n V_n + 1
\ea
and
\ba
N+ 2 P= \sum_n n V_n \, ,
\ea
the following relation between $L, V_n$ and $N$ is obtained, 
\ba
\label{relation1}
2 L= (2- N) + \sum_n (n-2) V_n \, .
\ea
For our case of interest, i.e.  the loop corrections in power spectrum, we have  $N=2$ so the above relation  simplifies to,
\ba
\label{relation2}
2 L=  \sum_n (n-2) V_n  \,  \quad \quad (N=2) \, .
\ea
The number of independent diagrams is the number of all consistent solutions of the above equation.  For example,  for one-loop corrections, there are only two solutions, {{\bf (a)}}: $V_3=2, V_4=0$ and {\bf (b)}: $V_3=0, V_4=1$. These correspond to two independent diagrams \cite{Firouzjahi:2023aum}, {{\bf (a)}}:  the diagram with two cubic  vertices and {{\bf (b)}}: the diagram  with a single quartic vertex.

Going to higher loop levels, the number of independent solution of 
Eq. (\ref{relation2}) increases rapidly. For example, for the case $L=2$, there exists  11 independent one-particle irreducible diagrams \cite{Firouzjahi:2024sce}. Calculating the contributions of all of these diagrams is a horrendous task. As a first step forward, the loop correction from a ``two scoop" diagram involving two vertices of quartic Hamiltonian were calculated in \cite{Firouzjahi:2024sce}. The conclusion was that the two-loop corrections in power spectrum scale like the square of the one-loop correction. 
It was shown that the loop correction can get of perturbative control in the  models with sharp transitions. 

The Feynman diagrams  at the $L$-loop order associated with the solution 
of Eq. (\ref{relation2}) appear in various shapes concerning the number of vertices $V_n$.  As the number of vertices increases for a given $L$,  the analysis 
becomes more complicated. This is because the diagrams with   multiple vertices require multiple nested in-in time integrals. For example, for $L=1$ 
the case of diagram {\bf (a)} listed above  involves two cubic vertices so one has to take a nested double time integral. As seen in \cite{Firouzjahi:2023aum}, the analysis for this diagram is far more complicated than the case {\bf (b)} listed above which involves a single quartic integral.

\begin{figure}[t]
\vspace{-1 cm}
	\centering
	\includegraphics[ width=0.9\linewidth]{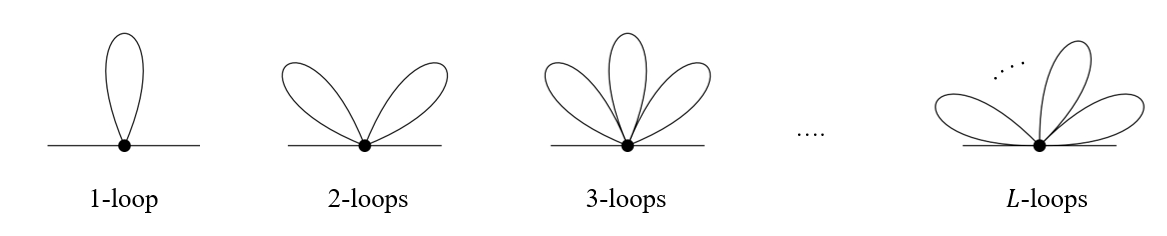}
	\vspace{0.5 cm}
	\caption{ The  one-vertex, one-particle irreducible 
	Feynman diagrams for various values of $L$.  For each value of $L$, we need the interaction Hamiltonian ${\bf H}_{2 L+2}$ to calculate the loop corrections.   }
\label{diagrams}
\end{figure}

With the above discussions in mind, in this work we consider a subset of Feynman diagram at $L$-loop which involves a single vertex, $V_n=1$.  This has the major advantage that one deals with a single in-in time integral. 
Setting $V_n=1$ in  Eq. (\ref{relation2}) fixes $n$, the order of 
self-interaction,  to be $n=2L+2$. Correspondingly, to calculate the $L$-loop corrections associated to this diagram, we need the interaction 
Hamiltonian   ${\bf H}_{2 L+2}$. 
For example, for $L=1$ we obtain $n=4$, corresponding to the quartic 
interaction ${\bf H}_{4}$,  for $L=2$ we obtain $n=6$, i.e. the sextic interaction ${\bf H}_{6}$ and so on. in Fig. \ref{diagrams}  we have plotted the shapes of these one-vertex diagrams for various values of $L$.

While considering the one-vertex diagram such as  in Fig. \ref{diagrams}
simplifies the analysis, but this brings no restrictions in understanding the amplitude of loop corrections. Indeed, as we shall see in the following analysis, for a given value of $L$,  the one-vertex diagram with the interaction Hamiltonian ${\bf H}_{2 L+2}$ has the most dominant contribution in loop corrections compared to all other multi-vertices diagrams. This is because the effects of the singular terms associated with the Dirac delta function $\delta(t-t_e)$ and its derivatives become 
more pronounced as one increases the order of interaction. Since for a given value of $L$ the interaction Hamiltonian ${\bf H}_{2 L+2}$ has the highest order among all interactions, then the loop corrections associated to this diagram is dominant compared to all other multi-vertices diagrams.

The general form of the interaction Hamiltonian, valid both during the bulk of USR as well as at the boundary $t=t_e$, is given  Eq.~\eqref{HI-eq2}. As we need the interaction Hamiltonian at the order $n=2L+2$, the corresponding Hamiltonian from the expansion of 
Eq.~\eqref{HI-eq2} is obtained to be
\ba
\label{H-int}
{\bf H}_{2L+2}=  \frac{M_P^2 H^2}{( 2 L)!} \int d^3 x\,   a(t)^3 \Big[ {\tilde\epsilon}^{(2 L)}   \dot \pi^2 + \epsilon^{(2 L)}  \partial_i\pi \partial^i \pi \Big]\, \pi^{2L}  \, ,
\ea
in which $X^{(k)}$ means the $k$-the order derivative, i.e. 
$X^{(k)}(z) \equiv \frac{d^{k}}{d z^{k}} X(z)$. In addition,  to simplify the notation, we have defined
\ba
{\tilde\epsilon}^{(2 L)} \equiv \epsilon(t)^2 \Big(\frac{1}{\epsilon(t)}\Big)^{(2 L)}\, .
\ea

Equipped with the above interaction Hamiltonian, we calculate the $L$-th order  loop corrections in the power spectrum $ \langle \calR_{\bfp_1}(\tau_0) \calR_{\bfp_2}(\tau_0) \rangle$ measured at the time of end of inflation $\tau=\tau_0$. In our notation, $\bfp_1$ and $\bfp_2$ represent the CMB scale momenta while the modes running in the loop are denoted by $\bfq$ or $\bfk$. Due to vast hierarchy involved between these two scales, we  assume $p_1, p_2 \ll q, k$ throughout the analysis.

An important point to note is that the interaction Hamiltonian 
(\ref{H-int}) is in terms of the Goldstone field $\pi$ while we are interested in power spectrum of $\calR$ which is non-linearly related to $\pi$ via 
Eq. (\ref{R-eq2}). However, at the time of end of inflation the system has reached to its attractor phase so time derivatives such as $\dot \pi, \ddot \pi, ...$  and $\dot H, \ddot H, ...$ etc. are all suppressed. Therefore, as long as the measurement is made at $\tau=\tau_0$, we can simply use the linear relation 
between $\calR$ and $\pi$ as follows,
\ba
\calR \rightarrow  - H \pi  \, , \quad \quad (\tau= \tau_0) \, .
\ea
Mathematically, this means that we can use the mode function of $\calR$ and 
$\pi$ interchangeably in the analysis so
\ba
\langle \calR_{\bfp_1}(\tau_0) \calR_{\bfp_2}(\tau_0) \rangle = 
H^2  \langle \pi_{\bfp_1}(\tau_0) \pi_{\bfp_2}(\tau_0) \rangle \, .
\ea
This should be compared with the case reviewed in section \ref{bispectrum} where the measurement is made at the 
end of USR, $\tau=\tau_e$, so the non-linear relation between $\calR$ and 
$\pi$ had to be taken into account.

To calculate the loop corrections in power spectrum, we use the perturbative in-in formalism  \cite{Weinberg:2005vy}, 
\ba
 \label{Dyson}
 &&\langle \calR_{\bfp_1}(\tau_0) \calR_{\bfp_2}(\tau_0) \rangle = 
 H^2  \langle \pi_{\bfp_1}(\tau_0) \pi_{\bfp_2}(\tau_0) \rangle \\
&&~~~~~~~~~~~~~~ = H^2
 \Big \langle \Big[ \bar {\mathrm{T}} \exp \Big( i \int_{-\infty}^{\tau_0} d \tau' {\bf H}_{2L+2} (\tau') \Big) \Big] \,  \pi_{\bfp_1}\pi_{\bfp_2}(\tau_0)  \, \Big[ \mathrm{T} \exp \Big( -i \int_{-\infty}^{\tau_0} d \tau' {\bf H}_{2L+2} (\tau') \Big) \Big]
 \Big \rangle \, , \nonumber
 \ea
in which the notations  $\mathrm{T}$ and $\bar {\mathrm{T}}$ indicate the usual time ordering and anti-time ordering respectively. 

For the single vertex diagram as depicted in Fig. \ref{diagrams}, there is only one time integral, yielding 
\begin{equation}
 \label{in-in-int}
 \langle \calR_{\bfp_{1}}(\tau_0) \calR_{\bfp_{2}}(\tau_0) \rangle
 = - 2 H^2   \mathrm{Im}  \int_{-\infty}^{\tau_0} d \tau   \big \langle {\bf H}_{2L+2} (\tau) \pi_{\bf p_1}(\tau_0) \pi_{\bf p_2}(\tau_0)  \big\rangle \, .
 \end{equation}
As discussed previously, if one considers diagrams involving multi vertices, then one has to deal with multiple nested time integrals which makes the analysis very complicated.

There are three different types of contributions in the integral (\ref{in-in-int}).
The first type is the contribution from the bulk of USR corresponding to 
$\tau_s< \tau < \tau_e$. The bulk contribution itself contains four distinct terms which we  calculate below. For the integrals in the bulk,  there is no contribution  from the discontinuities of $\eta$ and its derivatives which are encoded in 
$\delta (\tau-\tau_e)$ and its derivatives  so these integrals are relatively easy to handle. The second type of contribution concerns the boundary terms at $\tau=\tau_e$ which are associated with the jump in $\eta$. These are somewhat non-trivial to handle which we elaborate in more details below. The third type of the contribution concerns the time after the USR phase, 
$\tau_e < \tau< \tau_0$. However, since the mode function rapidly approaches its attractor value in the sharp transition limit, then one expects that this contribution to be subleading compared to the contribution from the bulk and the boundary.  This was specifically checked for $L=1$ in \cite{Firouzjahi:2023aum}. We have confirmed this conclusion numerically to many orders in perturbation theory   so we believe it to  hold for the general case as well. 
As a simple example to see this conclusion, consider the special case $h=-6$ which was considered in \cite{Kristiano:2022maq}. In this case,  $\epsilon(\tau)= \epsilon_e$ for $\tau> \tau_e$ so $\eta$ vanishes identically after $\tau=\tau_e$. Correspondingly, the interaction Hamiltonian  (\ref{H-int}) vanishes for $\tau> \tau_e$
and one has to consider ${\cal O}(\epsilon^2)$ terms which are discarded. 

With the above discussions in mind, in principle, there can be loop corrections induced from the boundary terms located at $\tau=\tau_i$, as there is a discontinuity in \(\eta\) at this point. However since the first stage of inflation is an SR phase, 
the mode functions at $\tau=\tau_i$ are not amplified—unlike during the USR stage. Therefore, one can safely neglect the loop corrections induced from the boundary terms at $\tau=\tau_i$ compared to loop corrections induced during the USR phase and from the boundary at $\tau=\tau_e$.

In the next two sections we calculate the loop corrections from the bulk of USR and the boundary $\tau=\tau_e$ in turn. 

\section{Loop Corrections from the Bulk}
\label{bulk}

We start with the contribution from the bulk, corresponding to $\tau_s < \tau < \tau_e$. Since in the bulk $\eta=-6$ is constant, then the Hamiltonian takes a simple form. More directly, from our non-perturbative formula  (\ref{HI-eq3}) 
the Hamiltonian ${\bf H}_{2 L+2}$ in the bulk  is easily obtained to be,
\ba
\label{H-bulk}
{\bf H}_{2 L+2}(\tau)= \frac{M_P^2 H^2}{( 2 L)!} a(\tau)^2 \epsilon(\tau)  (6 H)^{2 L} \int d^3 x\,  \Big( \pi'^2 + \partial_i \pi \partial^i \pi \Big)  \pi^{2 L}\, .
\ea
Alternatively, this can also be obtained from the general form Eq. (\ref{H-int}) as well. To see this, note that in the bulk $\epsilon(t) \propto a^{-6}$
so one can easily show that, 
\ba
\label{ep-tep}
{\tilde\epsilon}^{(2 L)}(t)= {\epsilon}^{(2 L)}(t)
= (6 H)^{2 L} \epsilon(\tau) \, .
\ea

Plugging the Hamiltonian (\ref{H-bulk}) into our master in-in integral Eq. (\ref{in-in-int}) we obtain, 
\ba
\label{in-in0}
 \langle \calR_{\bfp_{1}} \calR_{\bfp_{2}}(\tau_0) \rangle_{\mathrm{Bulk}}
 =\frac{-2 M_P^2 H^4}{(2L)!} ( 6 H)^{2 L} 
 \mathrm{Im} \int_{\tau_s}^{\tau_e} d \tau \epsilon(\tau) a^2 \int d^3 x  \Big\langle \pi(\tau)^{2 L} 
  \big(  \pi'^2 +  \big( \partial\pi\big)^2 \big)
\pi_{\bf p_1}\pi_{\bf p_2}(\tau_0)   \Big \rangle
\ea

The interaction Hamiltonian has two different types, the kinetic term and the
gradient term involving $\pi'^2$ and $\big( \partial\pi\big)^2$ in Eq. (\ref{in-in0}) respectively.  Let us first consider the kinetic term and for the moment let us concentrate on the spatial integral as follows,
\ba
{\cal I} \equiv  \int d^3 x  \Big\langle \pi(\tau)^{2 L} 
   \pi'(\tau)^2  \pi_{\bf p_1}\pi_{\bf p_2}(\tau_0)   \Big \rangle \, .
\ea
 Going to Fourier space, the above integral is cast into,
 \ba
 {\cal I} =\prod_i^{2L} \prod_j^2 \int \frac{d^3 \bfk_i}{(2 \pi)^3} \frac{d^3 \bfq_j}{(2 \pi)^3} \Big \langle 
 \hat\pi_{\bfk_1} \hat\pi_{\bfk_2}... \hat\pi_{\bfk_{2L}}(\tau)\, \,  \hat \pi'_{\bfq_1} \hat \pi'_{\bfq_2}(\tau)  \, \,  \hat\pi_{\bf p_1} \hat\pi_{\bf p_2}(\tau_0)
 \Big \rangle  
 \times ( 2 \pi)^3 \delta^3  \big(\sum_i {\bfk_i} + \sum_j {\bfq_j} \big)  \, .
 \nonumber
 \ea
 Here the quantities with a hat correspond to the quantum operator. 
There are four different contributions, depending on how $\hat\pi_{\bfp_1}$ and $\hat \pi_{\bfp_2}$  are contracted with $\hat\pi_{\bfk_i}$ and  $\hat\pi'_{\bfq_j}$.  However, one term which is obtained by the contraction of both $\hat\pi_{\bfp_1}$ and $\hat\pi_{\bfp_2}$ with the derivative terms $\hat \pi'_{\bfq_1}$ and $ \hat \pi'_{\bfq_2}(\tau)$ are suppressed. This is because this contraction involves two powers of  $\pi'_p(\tau_0)$ which will be suppressed in the soft limit $p\rightarrow 0$. Performing the calculations,  one can show that this contraction scales like $p^{-1}$. On the other hand, the remaining three contractions which are obtained either by contracting both $\hat\pi_{\bfp_1}$ and $\hat\pi_{\bfp_2}$ with $\hat\pi_{\bfk_i} $ or one of $\hat\pi_{\bfp}$ with 
$\hat\pi_{\bfk_i} $ and the remaining $\hat\pi_{\bfp}$  with $\hat \pi'_{\bfq_j}$, scale like $p^{-3}$ so these three contractions will dominate in the limit $p\rightarrow 0$ \cite{Firouzjahi:2023aum}.   

Now consider the term in  ${\cal I}$, denoted by $A_1$,
in  which $\hat \pi'_{\bfq_1}$ and $ \hat \pi'_{\bfq_2}(\tau)$ are contracted with each other. Neglecting the numerical prefactors,  performing all contractions and imposing the momentum conservations, this contribution has the following form,
\ba
(A_1): \int_{\tau_s}^{\tau_e} d \tau a(\tau)^2 \, \epsilon(\tau) \, 
 \mathrm{Im} \Big[ {\pi^*_p(\tau_0)}^2 \pi_p(\tau)^2 \Big] 
 \int \frac{d^3 q}{(2 \pi)^3}  \big|\pi'_q(\tau)\big|^2 
 \prod_i^{ L-1} \int \frac{d^3 k_i}{(2 \pi)^3}  \big|\pi_{k_i}(\tau)\big|^2 \, .
\ea
There are total $(2 L) (2 L-1)!!$ contractions which yield to the same result $A_1$ so this symmetry factor should be taken into account.  
Note that $(2L-1)!!$ means $(2 L-1) \times (2L-3)\times ...\times 3\times 1$. 

Consider the second term in the contractions of ${\cal I}$, denoted by $A_2$,  
in which  $\hat \pi'_{\bfq_j}$ do not contract with each other but contract to  
$\pi_{k_i}$. This term has the following form 
\ba
(A_2):
 \int_{\tau_s}^{\tau_e} d \tau a(\tau)^2 {\epsilon}(\tau) 
 \mathrm{Im} \Big[ {\pi^*_p(\tau_0)}^2 \pi_p(\tau)^2 \Big] 
\prod_j^{ 2}  \int \frac{d^3 q_j}{(2 \pi)^3} 
\mathrm{Re}\Big[ \pi'_{q_j} \pi^*_{q_j} (\tau) \Big]
 \prod_i^{ L-2} \int \frac{d^3 k_i}{(2 \pi)^3}  \big|\pi_{k_i}(\tau)\big|^2  .
\ea
This contribution has total $(2 L) (2L-2) (2L-1)!! $ symmetry factor. 

Finally, the third contribution denoted by $B$, is obtained by contracting one of $\hat\pi_{\bfp}$ with  $\hat\pi_{\bfk_i} $ and the other $\hat\pi_{\bfp}$  with $\hat \pi'_{\bfq_j}$. This term has the following form,
\ba
(B):
 \int_{\tau_s}^{\tau_e} d \tau a(\tau)^2  {\epsilon}(\tau)  
 \mathrm{Im} \Big[ {\pi^*_p(\tau_0)}^2 \pi_p(\tau) \pi'_p(\tau) \Big] 
  \int \frac{d^3 q}{(2 \pi)^3}  \mathrm{Re} \big[\pi'_{q}(\tau) \pi^*_q(\tau) \big] 
 \prod_i^{ L-1} \int \frac{d^3 k_i}{(2 \pi)^3}  \big|\pi_{k_i}(\tau)\big|^2 \, 
\ea
This contribution has the symmetry factor $2^2 (2L) (2 L-1)!!$.

Now consider the gradient contribution  from $(\partial \pi)^2$ term involving the following spatial integral,
\ba
{\cal J} \equiv \int d^3 x  \Big\langle \pi(\tau)^{2 L} 
  (\partial \pi)^2(\tau)   \pi_{\bf p_1}\pi_{\bf p_2}(\tau_0)   \Big \rangle \, .
\ea
As before, going to the Fourier space,  this yields to, 
 \ba
 {\cal J} =-\prod_i^{2L} \prod_j^2 \int \frac{d^3 \bfk_i}{(2 \pi)^3} \frac{d^3 \bfq_j}{(2 \pi)^3}  {\bfq_1}\cdot {\bfq_2}
 \Big \langle 
 \hat\pi_{\bfk_1} \hat\pi_{\bfk_2}... \hat\pi_{\bfk_{2L}}(\tau)  \hat \pi_{\bfq_1} \hat \pi_{\bfq_2}(\tau)   \hat\pi_{\bf p_1} \hat\pi_{\bf p_2}(\tau_0)
 \Big \rangle  
  ( 2 \pi)^3 \delta^3  \big(\sum_i {\bfk_i} + \sum_j {\bfq_j} \big) \nonumber
 \ea
The dominant contributions are those in which none of $ \hat \pi_{\bfq_1}$ contract with $\hat\pi_{\bf p}(\tau_0)$. This leave two possibilities. The first option is that $\hat \pi_{\bfq_1}$ contracts with each other. The second possibility is that $\hat \pi_{\bfq_1}$ does not contract with each other but contract with $\hat\pi_{\bfk_i}$. However, this contribution vanish dues to isotropy. More specifically, the latter contribution takes the following form 
\ba
\prod_m^{L-2} \int {d^3 \bfk_m}
\int d^3 k_1 \int d^3 k_2 \bfk_1 \cdot \bfk_2  \, \, 
{\cal F}(k_1, k_2, k_m;  \tau) \, .
\ea
However, this integral vanishes after integrating over 
 $\bfk_1$ and $\bfk_2$ due to isotropy of the background. 

Consequently, the only term associated with the gradient Hamiltonian, denoted by $\tilde A_1$, has the following form 
\ba
(\tilde {A_1}): 
\int_{\tau_s}^{\tau_e} d \tau a(\tau)^2 \, {\epsilon}(\tau) \, 
 \mathrm{Im} \Big[ {\pi^*_p(\tau_0)}^2 \pi_p(\tau)^2 \Big] 
 \int \frac{d^3 q}{(2 \pi)^3} q^2  \big|\pi_q(\tau)\big|^2 
 \prod_i^{ L-1} \int \frac{d^3 k_i}{(2 \pi)^3}  \big|\pi_{k_i}(\tau)\big|^2 \, .
\ea
The symmetry facto associated to this contribution is 
$(2 L) (2L-1)!!$. 

Combining the dominant contributions $A_1, A_2, B$ and $\tilde A_1$, the corrections in power spectrum from the integral in the bulk has the following form 
\begin{equation}
\label{eq:General Power}
\langle \mathcal{R}_{\textbf{p}_1}(\tau_0) \mathcal{R}_{\textbf{p}_1}(\tau_0)  \rangle = (2 \pi)^3 \delta^3(\textbf{p}_1 + \textbf{p}_2)  (A_1 + A_2 + B + \tilde{A}_1) \, .
\end{equation}

Collecting all the numerical factors,  the terms  
$A_1, A_2, B$ and $\tilde A_1$ are given as follows,

\ba
\label{A1-eq}
A_1 &=& \frac{2 M_P^2 H^4}{(2 L-2)!!}
\int_{\tau_s}^{\tau_e} d \tau a(\tau)^2 \,   {\tilde\epsilon}^{(2 L)}(\tau) \, 
 X(\tau)\, 
\Big(  \int \frac{d^3 q}{(2 \pi)^3}  \big|\pi'_q(\tau)\big|^2 \Big)
 \Big( \int \frac{d^3 k}{(2 \pi)^3}  \big|\pi_{k}(\tau)\big|^2 \Big)^{L-1}  ,\\
 \label{A2-eq}
A_2 &=& \frac{2 M_P^2 H^4}{(2 L-4)!!}
\int_{\tau_s}^{\tau_e} d \tau a^2  {\tilde\epsilon}^{(2 L)}(\tau)  
 X(\tau)
\Big(  \int \frac{d^3 q}{(2 \pi)^3} \mathrm{Re}\big[ \pi'_{q} \pi^*_{q} (\tau) \big] \Big)^2
 \Big( \int \frac{d^3 k}{(2 \pi)^3}  \big|\pi_{k}(\tau)\big|^2 \Big)^{L-2 } \\
 \label{B-eq}
B &=& \frac{8 M_P^2 H^4}{(2 L-2)!!}
\int_{\tau_s}^{\tau_e} d \tau a^2 \, {\tilde\epsilon}^{(2 L)}(\tau)  
 Y(\tau)
\Big(  \int \frac{d^3 q}{(2 \pi)^3} \mathrm{Re}\big[ \pi'_{q} \pi^*_{q} (\tau) \big] \Big)
 \Big( \int \frac{d^3 k}{(2 \pi)^3}  \big|\pi_{k}(\tau)\big|^2 \Big)^{L-2} ,
\ea
and
\ba
\label{tildeA1-eq}
\tilde{A_1} &=& \frac{2 M_P^2 H^4}{(2 L-2)!!}
\int_{\tau_s}^{\tau_e} d \tau a^2 \, {\epsilon}^{(2 L)}(\tau) \, 
 X(\tau)
\Big(  \int \frac{d^3 q}{(2 \pi)^3} q^2  \big|\pi_q(\tau)\big|^2 \Big)
 \Big( \int \frac{d^3 k}{(2 \pi)^3}  \big|\pi_{k}(\tau)\big|^2 \Big)^{L-1} \, ,
\ea
in which to simplify the expression, we have defined $X(\tau)$ and $Y(\tau)$
as follows,
\ba
X(\tau) \equiv \mathrm{Im} \Big[ {\pi^*_p(\tau_0)}^2 \pi_p(\tau)^2 \Big] \, ,
\quad \quad
Y(\tau)\equiv \mathrm{Im} \Big[ {\pi^*_p(\tau_0)}^2 \pi_p(\tau) \pi'_p(\tau) \Big] 
\, .
\ea
Furthermore, we have kept ${\tilde\epsilon}^{(2 L)}$  and 
${\epsilon}^{(2 L)}$ in the above integrals for similar analysis for the boundary terms in next section. But, here, in the bulk,  ${\tilde\epsilon}^{(2 L)}$ and  ${\epsilon}^{(2 L)}$ are given in Eq. (\ref{ep-tep}).

The integrations above can be performed analytically once we consider the structure of the background solutions and the mode functions during the USR period. We have presented the mode functions in all three phases 
in Appendix \ref{mod-functions}. In particular, during the USR phase 
the mode function on superhorizon scales  satisfies the 
relation $\pi_k'(\tau)= -\frac{3}{\tau} \pi_k(\tau)$ which can be used in the above integrals.   

On the other hand, using the form of the mode function, the quantities 
$X(\tau)$ and $Y(\tau)$ are obtained to be \cite{Firouzjahi:2023aum}, 
\ba
X(\tau)= \frac{H^2}{24 M_P^2 \epsilon_i^2 p^3} \frac{\tau_s^6}{\tau_e^3 \tau^3}
\Big(  \frac{6-h}{h} \tau^3 + \tau_e^3    \Big)\, ,
\ea
and
\ba
Y(\tau)= \frac{-H^4 \tau_s^6}{16 M_P^4 \epsilon_i^2 \tau^4 p^3} \, .
\ea
Finally, using the evolution of the mode function we can calculate the following common integral as well,
\ba
\int \frac{d^3 k}{(2 \pi)^3}  \big|\pi_{k}(\tau)\big|^2 = 
\Big( \frac{\tau_e}{\tau} \Big)^6  \int \frac{d^3 k}{(2 \pi)^3}  
\big|\pi_{k}(\tau_e)\big|^2 \, .
\ea

The integral $d^3 k$ above should be over all momenta. Let us define $k_i$ and $k_e$ as the modes which leave the horizon at $\tau=\tau_i$ and $\tau=\tau_e$ respectively.  There are three types of contributions: the IR contribution corresponding to $0< k < k_i$, the USR modes contributions with $k_i \leq k \leq k_e$, and the UV contributions for $k > k_e$.  As the IR modes are not amplified, we do not expect their contributions to be important, and in any case, they can be regularized by imposing a large box cutoff. The USR modes are amplified, so we calculate their contributions in more detail below. Finally, we expect that the contributions of the  UV modes will be divergent as in standard QFT analysis, which requires regularization and renormalization.

With the above discussions in mind, and considering the USR modes in the range $k_i \leq k \leq k_e$,  the integral above is simply related to the power spectrum at the end of USR via, 
\ba
\int \frac{d^3 k}{(2 \pi)^3}  \big|\pi_{k}(\tau_e)\big|^2 = \frac{e^{6 \Delta N}}{4 M_P^2 \epsilon_i } \int_{k_i}^{k_e}  \frac{d k}{2 \pi^2 k} = \frac{\calP_{\mathrm{cmb}}}{H^2} ( \Delta N e^{6 \Delta N}) \, ,
\ea
in which $\calP_{\mathrm{cmb}}$ is the power spectrum on CMB scale,
\ba
\calP_{\mathrm{cmb}} = \frac{H^2}{8 \pi^2 \epsilon_i M_P^2 } \, .
\ea

With the above results in hands, we can calculate $A_1, A_2, B$ and $\tilde A_1$. Starting with $A_1$ and defining the dimensionless loop correction by
\ba
\Delta \calP_{A_1}\equiv  \frac{p^3}{2 \pi^2} A_1 \, ,
\ea
we obtain the following expression for the fractional loop correction from the $A_1$ term,
\ba
\label{A1-loop}
\frac{\Delta \calP_{A_1} }{\calP_{\mathrm{cmb}}}\Big|_{L-\mathrm{loop}} = \frac{(12 L-h)}{(2 L-1)h}
\frac{1}{\Gamma(L+1)} \Big(18 \Delta N \calP_e \Big)^L \, ,
\ea
in which $\calP_e$ is the power spectrum at the end of USR phase, 
\ba
\label{Pe}
\calP_e = \frac{H^2}{8 \pi^2 M_P^2 \epsilon_e} =  
\calP_{\mathrm{cmb}} e^{6 \Delta N}\, .
\ea
In obtaining the above result, the following mathematical identity is used,
\ba
( 2 k)!! = 2^{ k} \Gamma(k+1) \, .
\ea

To calculate other terms, we note that there are similarities in the   
structure of integrals in Eqs. (\ref{A1-eq}), (\ref{A2-eq}), (\ref{B-eq}) and
(\ref{tildeA1-eq}). In particular, one can check that the following relations hold,
\ba
\label{A2-term}
\Delta \calP_{A_2}\Big|_{L-\mathrm{loop}} = 2 (L-1) \Delta \calP_{A_1}\Big|_{L-\mathrm{loop}} \, ,
\ea
and, 
\ba
\Delta \calP_{B}\Big|_{L-\mathrm{loop}} =  \frac{2 (2 L-1)}{12 L- h} h \Delta \calP_{A_1}\Big|_{L-\mathrm{loop}} \, .
\ea

Finally, to calculate the contribution of the  gradient term $\tilde A_1$, noting that in the bulk $\epsilon^{(2 L)}= \tilde \epsilon^{(2L)}$,  we obtain,
\ba
\Delta \calP_{\tilde A_1}\Big|_{L-\mathrm{loop}}= \frac{1}{\Delta N}\frac{L (2 L-1) (12 L-h-4)}{2 (6 L-5) (3 L-1) ( 12L-h)}  \Delta \calP_{A_1}
\Big|_{L-\mathrm{loop}}\, .
\ea
In particular, note that the contribution of the $\tilde A_1$  is accompanied by a factor $1/\Delta N$ as observed in \cite{Firouzjahi:2023aum}. 
In the limit $L\gg 1$, the dominant contribution amongst $A_1, A_2, B$ and $\tilde A_1$ comes from the $A_2$ term as can be seen in Eq. (\ref{A2-term}). 

Combining all four contributions, the fractional loop corrections from the bulk terms are given by,
\ba
\label{Bulk-cont}
\frac{\Delta \calP_{\mathrm{Bulk}} }{\calP_{\mathrm{cmb}}}\Big|_{L-\mathrm{loop}}=
R_{\mathrm{B}}(L)  \big( \Delta N \calP_e \big)^L \, ,
\ea
in which the numerical prefactor  $R_{\mathrm{B}}$ is defined via, 
\ba
R_{\mathrm{B}}(L)  \equiv \Big[ {12 L+ h} + \frac{1}{\Delta N}\frac{L  (12 L-h-4)}{2 (6 L-5) (3 L-1) } \Big] 
\frac{18^L}{h \Gamma(L+1)} \, .
\ea
Beside the numerical factor $R_{\mathrm{B}}(L)$, it is quite interesting and reassuring that the loop correction at the $L$-th order appears 
as $\big( \Delta N \calP_e \big)^L$. This was observed for the two-loop 
case in \cite{Firouzjahi:2024sce}. 

Looking at the  behaviour of  $R_{\mathrm{B}}(L) $ one can show that it falls off with large $L$. More specifically, 
\ba
\label{RB-assym.}
R_{\mathrm{B}}(L) \rightarrow \frac{6 \sqrt{2 L} }{h \sqrt{\pi}} \big( \frac{18 e}{L}\big)^L  \, , \quad \quad (L\gg 1) \, .
\ea
This suggests that the total contribution in the bulk from all loops should be converging. Interestingly, we are able to do the resummation over all loop contributions in the bulk, obtaining non-perturbative results. More specifically, the contributions of the kinetic terms $A_1+A_2+ B$ is summed to, 
\footnote{We use Maple and Mathematica softwares  to perform the analytical and numerical analysis.}
\ba
\label{resum-K}
\frac{\Delta \calP^{\mathrm{kin}}_{\mathrm{Bulk}} }{\calP_{\mathrm{cmb}}}
\Big|_{\mathrm{resummed}}
=-1 + {e^{18 \Delta N \calP_e} } \Big( 1+ \frac{216}{h} \Delta N \calP_e \Big) \, .
\ea
The resummation over the gradient term $\tilde A_1$ yields, 
\ba
\label{resum-G}
\frac{\Delta \calP^{\mathrm{grad}}_{\mathrm{Bulk}} }{\calP_{\mathrm{cmb}}}\Big|_{\mathrm{resummed}}
= -\Big( \frac{9 (h-8)}{2 h }\, \calP_e\Big)\,    {_3}\mathrm{F}_3\, \Big(\Big[\frac{1}{6}, \frac{2}{3}, \frac{20-h}{12} \Big],  \Big[\frac{7}{6}, \frac{5}{3}, \frac{8-h}{12} \Big], 18 \Delta N \calP_e \Big) \, .
\ea
in which ${_3}\mathrm{F}_3$ is the generalized hypergeometric function.

The total resummed loop corrections from the bulk and boundary from the sum of the above two contributions is given by,
\ba
\label{resum-bulk}
\frac{\Delta \calP^{\mathrm{tot}}_{\mathrm{Bulk}} }{\calP_{\mathrm{cmb}}}
\Big|_{\mathrm{resummed}}
&=&-1 + {e^{18 \Delta N \calP_e} } \Big( 1+ \frac{216}{h} \Delta N \calP_e \Big) \nonumber\\
&-& \Big( \frac{9 (h-8)}{2 h }\, \calP_e\Big)\,    {_3}\mathrm{F}_3\,   \Big(\Big[\frac{1}{6}, \frac{2}{3}, \frac{20-h}{12} \Big],  \Big[\frac{7}{6}, \frac{5}{3}, \frac{8-h}{12} \Big], 18 \Delta N \calP_e \Big)\, .
\ea
The fact that the loop correction can be resummed as given in Eq. (\ref{resum-bulk}) is quite interesting and perhaps surprising. This result may be compared with the case of IR loops in finite temperature field theory where the resummation over all IR loops can be performed as well \cite{Kapusta:2007xjq}. More specifically, the dS spacetime can be viewed as a thermal background with the Hawking temperature 
$\frac{H}{2\pi}$ so the resummation here may have some technical as well as conceptual similarities with the loops resummation in finite temperature field theory. 

\begin{figure}[t!]
	\centering
	\includegraphics[ width=0.4\linewidth]{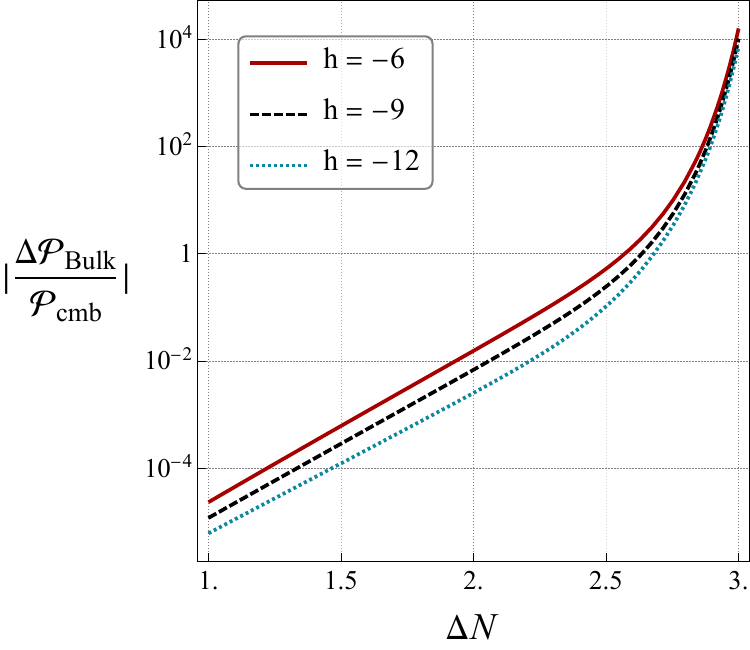}
	\hspace{+1.5 cm}\includegraphics[ width=0.4
	\linewidth]{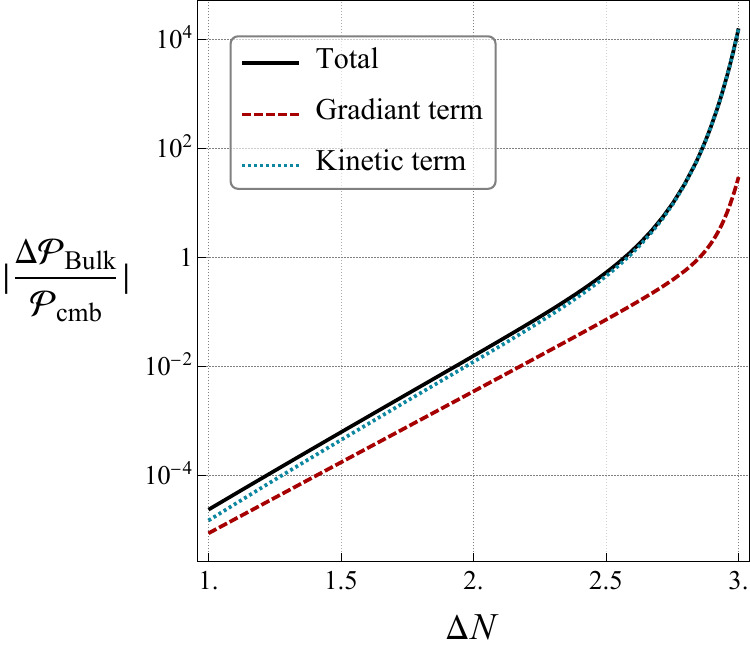}
	\caption{ {\bf Left:} The total resummed bulk loop corrections  from  Eq. (\ref{resum-bulk}). The curves from top to bottom correspond to $h=-6, h=-9$ and $h=-12$ respectively.  
	{\bf Right:} The comparison of the resummed bulk contributions for $h=-6$: the kinetic term Eq. (\ref{resum-K}) (dotted curve), the gradient term Eq. (\ref{resum-G}) (dashed curve) and the total contribution (\ref{resum-bulk}) (solid curve).  }
\label{totbulk}
\end{figure}


The hypergeometric function in Eq. (\ref{resum-bulk}) is not easy to examine.  But one can check that for small $\Delta N$, the two contributions 
(\ref{resum-G}) and (\ref{resum-K}) are comparable while for 
large enough $\Delta N$, say $\Delta N \gtrsim 2.5$,  
the dominant contribution in resummed loop correction comes  from the kinetic terms (\ref{resum-K}) which scales like 
${e^{18 \Delta N \calP_e} }$. Noting also that $\calP_e = \calP_{\mathrm{cmb}} e^{6 \Delta N}$, we conclude that if the duration of USR phase $\Delta N$ is long enough then the resummed loop correction soon gets out of perturbative control.  In the left panel of Fig. \ref{totbulk}, we have presented the resummed total bulk loop corrections given in   Eq. (\ref{resum-bulk}).  From this plot see that the loop corrections indeed get out of perturbative control for 
$\Delta N > 2.5$. In the right panel of this figure we have compared 
the two contributions (\ref{resum-G}) and (\ref{resum-K}) for $h=-6$. 

Although the non-perturbative result Eq. (\ref{resum-bulk}) for the loop corrections from the bulk of USR is an interesting result of its own right, but to complete our investigations of the total loop corrections, we should also calculate the loop corrections induced from the boundary at $\tau=\tau_e$.  It turns out that the loop corrections from the boundary are significantly more important than the corrections from the bulk 
and, indeed, the loop corrections do not converge for large value of 
$L$ as we will show below.

\section{Loop Corrections  from the Boundary of USR}
\label{boundary}

Here we calculate the  loop corrections from the boundary of USR at $t=t_e$. As discussed before, these contributions originate from the jump in $\eta$ which appears in $\dot \eta, \ddot \eta, \dddot \eta, \dot \eta^2$ etc which are encoded in the Dirac delta function $\delta (t- t_e)$ and its derivatives. These localized terms appear in $\epsilon^{(2 L)}$ and $ \tilde \epsilon^{(2L)}$ which can be seen in the expressions of 
$H^{(n)}$ in Eqs. (\ref{H3-eq})-(\ref{H4-eq}) as well.  

There are complexities in dealing with the boundary terms. Not only the higher derivatives of $\eta$ such as $\ddot \eta, \dddot {\eta}$ etc appear, but terms like $\dot \eta^2, \dot \eta^3$ and higher powers appear as well. These terms  behave like the powers of the Dirac delta function which are ill-defined mathematically.  

To handle this issue, it is convenient to work with the number of e-folds $N$ as the clock and express $\epsilon$ near the the boundary in terms of $N$. Without loss of generality, we set the point of boundary to be 
$N=0$ so the bulk of USR corresponds to $N<0$ while the period after the USR corresponds to $N>0$. Following the prescription of the evolution of 
$\eta$ given in section \ref{SETUP}, we parameterize it as follows,
\ba
\label{eta-N}
\eta(N) = \eta_1 \Theta(-N) + \eta_2 \Theta(N) \, ,
\ea
in which $\eta_1$ is the value of $\eta$ during the USR phase which is 
$\eta_1=-6$, while $\eta_2$ represents its value immediately after the transition which is $\eta_2= -(6+h)$ as given in Eq. (\ref{eta-jump1}). In particular, note the jump in $\eta$ from $\eta_1$ to $\eta_2$ with $\Delta \eta = -h$.  Here $\Theta(N)$ is the Heaviside step function which satisfies the following relation,
\ba
\frac{d \Theta(N)}{d N} = \delta (N)\, .
\ea 
Using the form of $\eta(N)$ from Eq. (\ref{eta-N}), the evolution of $\epsilon$ is given by,
\ba
\label{epsilon-N}
\epsilon(N) = \epsilon_1(N)\Theta(-N) + \epsilon_2(N) \Theta(N)\, ,
\ea
in which 
\ba
\epsilon_1(N)=  \epsilon_e e^{\eta_1 N}=\epsilon_e e^{-6 N}\, , \quad \quad
\epsilon_2(N)=  \epsilon_e e^{\eta_2 N}= \epsilon_e e^{-(6+h) N}\, .
\ea

With the above form of $\epsilon(N)$, the quantities 
$\epsilon^{(2 L)}$ is expressed in terms of derivatives of $N$ as follows
\ba
\label{ep-N}
\epsilon^{(2 L)}(t) = \frac{d^{2L} \epsilon(t)}{dt^{2L}}  &=& 
H^{2 L} \epsilon^{(2 L)}(N) 
+ {\cal O} (\epsilon^2) \, \nonumber\\
&=& H^{2 L}\Big(  \epsilon_1(N)\Theta(-N) + \epsilon_2(N) \Theta(N) \Big)^{(2 L)} + {\cal O} (\epsilon^2) \, ,
\ea
in which we have discarded terms of ${\cal O} (\epsilon^2)$ which comes from derivatives of $H$ in passing from $t$ to $N$. 

Similarly, for $ \tilde \epsilon^{(2L)}$ we obtain, 
\ba
\label{tep-N}
\tilde{\epsilon}^{(2 L)}(t)  = H^{2 L}  
\Big(  \epsilon_1(N)\Theta(-N) + \epsilon_2(N) \Theta(N) \Big)^2 
\Big(  \epsilon_1^{-1}(N) \Theta(-N) + \epsilon_2^{-1}(N) \Theta(N)  \Big)^{(2L)}  + {\cal O} (\epsilon^2) \,
\ea

Despite the similarities between $\epsilon^{(2 L)}$ and $ \tilde \epsilon^{(2L)}$ there are  subtle differences between them which play important roles at the boundary  $N=0$.  Roughly speaking,  $\epsilon^{(2L)}$ contains terms like 
$\delta (N)$, $\delta^{(1)}(N)$, $\delta^{(2)}(N)$ and so on. However, the quantity $ \tilde \epsilon^{(2L)}$ 
contains terms like $\Theta(N) \delta (N), \Theta(N) \delta^{(1)}(N)$ etc. which make the process of integration by parts non-trivial.

To calculate the loop corrections from boundary terms, let us start with the gradient term $\tilde A_1$ which involves the quantity  $ \epsilon^{(2L)}$ which is easier to handle. The functional form of $\tilde A_1$ is given in 
Eq. (\ref{tildeA1-eq}) except that now we should perform the time integral across the point of boundary $\tau_e^-< \tau <\tau_e^+$ corresponding to 
$0^-< N<0^+$.  Absorbing all the numerical factors and the 
smooth time-dependent functions  involved in $\tilde A_1$ 
into a smooth function $f_{\tilde A_1}(N)$, the integral for $\tilde A_1$ across the boundary has the following form,
\ba
\label{tildeA1-int}
{I_{\tilde A_1}} \equiv 
\int_{0^-}^{0^+} dN f_{\tilde A_1}(N) \epsilon^{(2 L)}(N)= 
\int_{0^-}^{0^+} dN f_{\tilde A_1}(N)  \big( \frac{d^{2L} }{dN^{2L}} \big) \epsilon(N) \, .
\ea 
The above integral can be taken via integration by parts. At the first step of 
integration by parts, this yields,
\ba
{I_{\tilde A_1}}= \Big[ f_{\tilde A_1}(N) \epsilon^{(2 L-1)}(N)\Big]_{0^-}^{0^+} -
\int_{0^-}^{0^+} dN f^{(1)}_{\tilde A_1}(N)  \, \epsilon^{(2 L-1)}(N) \, .
\ea
This means that after one step of integration by parts the power of derivative of $\epsilon(N)$ is reduced by one order but the derivative is transferred into the function $f_{\tilde A_1}(N)$ plus a term which measure the jump of the integrand across the boundary. Using the specific form of $\epsilon(N)$ from 
Eq. (\ref{epsilon-N}),  the first term above is 
\ba
\Big[ f_{\tilde A_1}(N) \epsilon^{(2 L-1)}(N)\Big]_{0^-}^{0^+}= 
\epsilon_e  f_{\tilde A_1}(0)\Big( \eta_2^{2 L-1} - \eta_1^{2 L-1} \Big) \equiv 
\epsilon_e f_{\tilde A_1}(0) \Delta \eta^{2 L-1} \, .
\ea
Continuing the integration by parts to second step, we obtain
\ba
{I_{\tilde A_1}}= 
\epsilon_e  f_{\tilde A_1}(0) \Delta \eta^{2 L-1} - \Big[ f^{(1)}_{\tilde A_1}(N) \epsilon^{(2 L-2)}(N)\Big]_{0^-}^{0^+} + \int_{0^-}^{0^+} dN f^{(2)}_{\tilde A_1}(N)  \, \epsilon^{(2 L-2)}(N) \, .
\ea
Performing the integration by parts in successive steps, we end up with the last integral which will be of the form,  
\ba
\int_{0^-}^{0^+} dN f_{\tilde A_1}^{(2 L)}  \, \epsilon(N) \, ,
\ea
which vanishes across the boundary. All that is left is the jumps in the integrands during these successive steps, yielding
\ba
\label{in-sum}
{I_{\tilde A_1}} &=& \epsilon_e\sum_{k=1}^{2 L} (-1)^{k-1}f_{\tilde A_1}^{( k-1)} (0)\Delta \eta^{2 L-k} \, ,
\nonumber\\ &=& - \epsilon_e \sum_{k=1}^{2 L} f_{\tilde A_1}^{( k-1)}(0) \Big(  (6+h)^{2L-k} - 6^{2L-k}
\Big) \, .
\ea 

Now let us calculate the boundary terms originated from the kinetic interactions $A_1, A_2$ and $B$ which involve the quantity 
$\tilde \epsilon^{(2 L)}(t)$. As mentioned above and 
elaborated in Appendix \ref{eps-relation}, the case $\tilde \epsilon^{(2 L)}(t)$ is more subtle than the case of 
$\epsilon^{(2 L)}(t)$. However, we have shown in Appendix \ref{eps-relation} 
that at the boundary the following identification holds,
\ba
\label{prescription}
\epsilon^{(2 L)}  \cong  - \tilde \epsilon^{(2 L)} \, ,  \quad \quad (N=0) \, .
\ea
It is understood that the above relation holds only at the point of boundary $N=0$ and is not an identity in the whole region. 

With the identification (\ref{prescription}), the integrals in the boundary for the terms $A_1, A_2$ and $B$ can be performed similar to the case 
$\tilde{A_1}$ obtained above. For example, consider the  case $A_1$.  Combining all the numerical factors and the smooth functions involved in the integral of $A_1$  into a smooth function $ f_{ A_1}(N)$, we have
\ba
{I_{ A_1}} \equiv 
\int_{0^-}^{0^+} dN f_{ A_1}(N) \tilde \epsilon^{(2 L)}(N)= \epsilon_e
\sum_{k=1}^{2 L} f_{ A_1}^{( 2 k)}(0) \Big(  (6+h)^{2L-k} - 6^{2L-k}
\Big) \, .
\ea 

Therefore, to calculate the boundary contributions, all that is left is to construct the smooth function $f_{ A_1}(N)$ or $f_{\tilde  A_1}(N)$ as defined above 
and plug them into the above sum.  For example, from Eq. (\ref{A1-eq}) 
the function $f_{ A_1}(N)$, including all the numerical factors,  is obtained to be, 
\ba
f_{ A_1}(N) = -\frac{3 H^2 \Big(h e^{6(L+1)N} + (6-h)  e^{(6L+3)N}   \Big) }{2^{L+1} M_P^2 \epsilon_i h \Gamma(L) \, p^3 } \calP_{\mathrm{cmb}}^L \, .
\ea

Calculating each term in the boundary as outlined above, the fractional  loop corrections  are obtained to be, 
\ba
\label{A1-te}
\frac{\Delta \calP_{A_1}}{\calP_{\mathrm{cmb}}}&=& \frac{6 \big( \Delta N \calP_e\big)^L}{2^L \Gamma(L) }\, 
\sum_{k=1}^{2 L } \Big(  (6+h)^{2L-k} - 6^{2L-k}
\Big) \Big[   (6+ 6 L)^{k-1}+ \frac{(6-h)}{h} ( 6L+ 3)^{k-1} \Big]   \, , \\
\label{A2-te}
\frac{\Delta \calP_{A_2}}{\calP_{\mathrm{cmb}}} &=& 2 (L-1)  \frac{\Delta \calP_{A_1}}{\calP_{\mathrm{cmb}}} \, , \\
\frac{\Delta \calP_{B}}{\calP_{\mathrm{cmb}}} &=& \frac{12 \big( \Delta N \calP_e\big)^L}{2^L \Gamma(L) }
\sum_{k=1}^{2 L } \Big(  (6+h)^{2L-k} - 6^{2L-k} \Big)   (6+ 6 L)^{k-1}  \, ,\\
\label{tA1-te}
\frac{\Delta \calP_{\tilde{A_1}}}{\calP_{\mathrm{cmb}}} &=& 
\frac{-\big( \Delta N \calP_e\big)^L}{3\Delta N  2^L \Gamma(L) }
\sum_{k=1}^{2 L } \Big(  (6+h)^{2L-k} - 6^{2L-k} \Big) 
\Big[  (4+ 6 L)^{k-1}+ \frac{(6-h)}{h} ( 6L+ 1)^{k-1} \Big]  .
\ea
The sums over $k$ in the above expressions can be performed which yield to long expressions which we present in Appendix \ref{sum-L}.

Combining the contributions of all four terms, the fractional loop correction 
from the boundary terms is summarized as follows, 
\ba
\label{boundary-cont}
\frac{\Delta \calP_{\mathrm{boundary}} }{\calP_{\mathrm{cmb}}}\Big|_{L-\mathrm{loop}}=
R_{\mathrm{b}}(L)  \big( \Delta N \calP_e \big)^L \, ,
\ea
in which the dimensionless parameter $R_{\mathrm{b}}(L)$ is obtained by combining the sums over $k$ as given in Appendix \ref{sum-L}. 
 
As in the case of bulk contributions, it is interesting and reassuring that the loop correction scales with power spectrum 
like $\big( \Delta N \calP_e \big)^L$. However, there are significant differences between the contributions of the bulk and boundary which are encoded in  dimensionless quantities $R_{\mathrm{B}}(L)$ and
$R_{\mathrm{b}}(L)$ respectively.  As can be seen from Eq. (\ref{RB-assym.}), $R_{\mathrm{B}}(L)$ falls of rapidly for large $L$. However, this does not happen for $R_{\mathrm{b}}(L)$. Looking at the  asymptotic behaviour of $R_{\mathrm{b}}(L)$ for  $L\gg 1$ in Appendix \ref{sum-L}, we obtain, 
\ba
R_{\mathrm{b}}(L)  \rightarrow \frac{\sqrt{2}(6+ (e-1)h) }{6 \sqrt{\pi}}
\Big ( 18^L e^{L+1} L^{L-\frac{1}{2}} \Big)  \sim ( 18 L)^L \,  
\quad \quad (L\gg 1) \, .
\ea
Correspondingly, the fractional loop correction in power spectrum for large values of $L$ scales like, 
\ba
\label{boundary-large}
\frac{\Delta \calP_{\mathrm{boundary}} }{\calP_{\mathrm{cmb}}}\Big|_{L-\mathrm{loop}}
\sim \big(18 L  \Delta N \calP_e \big)^L \,, \quad \quad  (L\gg 1) \, .
\ea
This demonstrates that the loop corrections induced from the boundary terms quickly get out of perturbative control and do not converge.  

The reason for the rapid growth of the loop corrections from the boundary for  $L\gg 1$ is that the boundary terms involve the singular contribution
$\delta (\tau-\tau_e)$ and its powers which are encoded in variables $ \epsilon^{(2 L)}(t)$ and $\tilde \epsilon^{(2 L)}(t)$.   We have employed integrations by parts $2L$ times to get rid of the singularities associated with $\delta (\tau-\tau_e)$ and its derivatives. This in turn induced the sums such as in Eq. (\ref{in-sum}) which  grow rapidly for large values of $L$. While the dependence of the loop corrections in terms of $\calP_e$ is the same for both bulk and boundary, scaling like $\big( \Delta N \calP_e \big)^L$, but it is the roles of large numbers encoded in $R_{\mathrm{b}}(L)$ which control the rapid growth of the loop correction induced from the boundary terms at $\tau=\tau_e$.  In the left panel of Fig. \ref{totboundary} we have compared the loop correction from the boundary with that of the bulk. From this plot it is seen that indeed the loop correction by the boundary grows rapidly for large $L$ while the contribution of the bulk falls off. 

\section{Total Loop Corrections}
\label{total}

Finally, combining the contributions form the boundary and the bulk, 
Eqs. (\ref{boundary-cont}) and  (\ref{Bulk-cont}), the total $L$-loop correction is obtained to be,
\ba
\label{total-loop}
\frac{\Delta \calP_{\mathrm{total}} }{\calP_{\mathrm{cmb}}}\Big|_{L-\mathrm{loop}}&=&
\Big( R_{\mathrm{b}}(L) + R_{\mathrm{B}}(L) \Big)  \big( \Delta N e^{6 \Delta N}  \calP_{\mathrm{cmb}} \big)^L \\
 \label{approximation}
 &\sim& \Big( 18 L  \Delta N e^{6 \Delta N} \calP_{\mathrm{cmb}}\Big)^L \, 
 \quad \quad (L \gg 1) \, .
\ea

\begin{figure}[t]
	\centering
	\includegraphics[ width=0.4\linewidth]{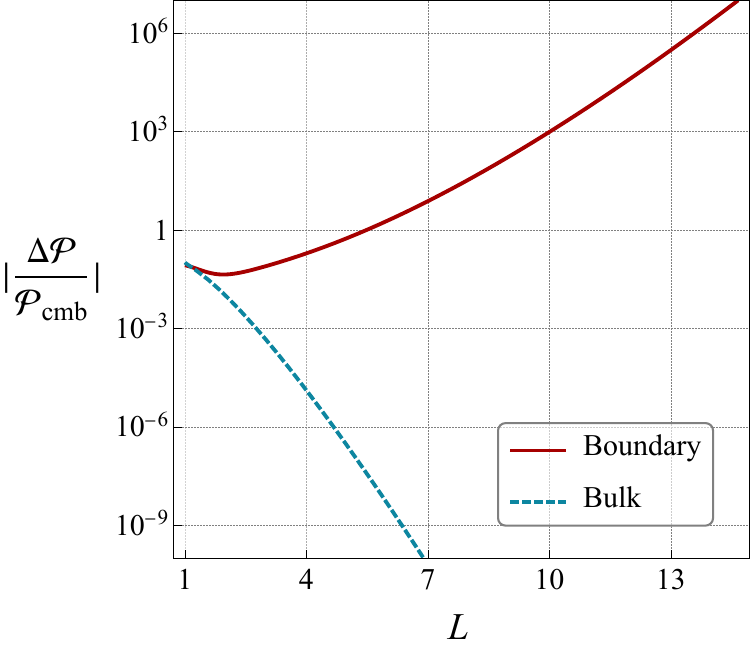}
	\hspace{+1.5 cm}\includegraphics[ width=0.4\linewidth]{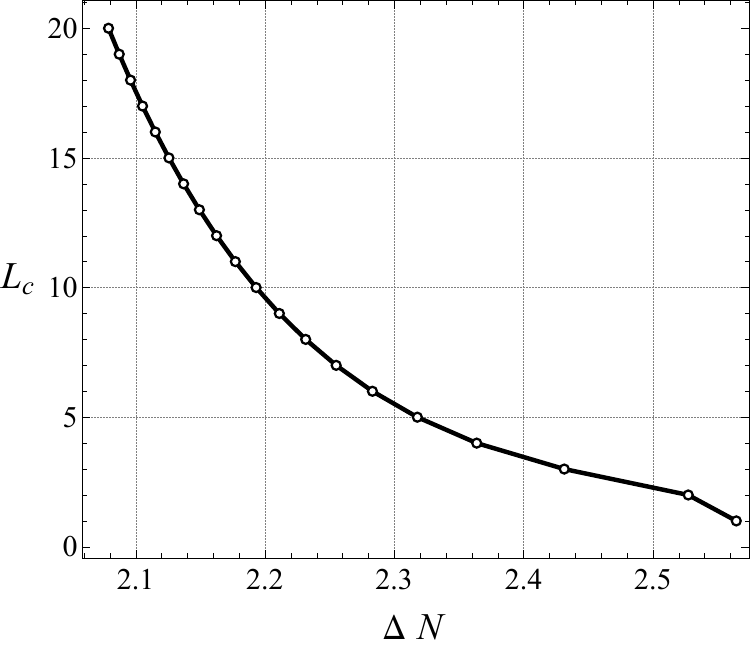}
	\caption{ {\bf Left:} The comparison of the contributions from the bulk Eq. (\ref{Bulk-cont}) (dashed curve) and the boundary Eq. (\ref{boundary-cont}) (solid curve) for $\Delta N=2.3$ and $h=-6$.  We see that for large $L$ the contribution of bulk falls off while that of boundary grows rapidly. {\bf Right:}  $L_c$ as a function of $\Delta N$ for $h=-6$.  The larger is $\Delta N$, the smaller is $L_c$. }
\label{totboundary}
\end{figure}


 There are  a number of  comments in order. First,  Eq. (\ref{total-loop}) indicates the strong sensitivity to the duration of USR phase $\Delta N$. Second, 
 as discussed before, the dominant contributions in $L$-loop correction come from the boundary term in which $R_{\mathrm{b}}(L) \gg R_{\mathrm{B}}(L)$ for $L\gg1$.  Third,  as can be seen from Eq. (\ref{approximation}), for any given value of $\Delta N$, there is a critical value  $L= L_c$ beyond which the loop corrections will get out of perturbative control. The value of $L_c$ is roughly given by
 \ba
 L_c \simeq \Big( 18  \Delta N e^{6 \Delta N} \calP_{\mathrm{cmb}}\Big)^{-1}\, .
 \ea 
 The larger is $\Delta N$, the smaller is this threshold value of $L$.  For example, in conventional USR setup considered for PBHs formation, one requires  $\Delta N \simeq 2.5$ e-folds to raise the power spectrum by seven order of magnitude to obtain $\calP_e \sim 10^{-2}$. This value of $\Delta N$ yields  the critical value $L_c \simeq 3.4$ so the loop corrections in this setup will get out of control at four-loop order. 
 This is a very interesting result. This result sheds new light on the conclusion of \cite{Kristiano:2022maq}, indicating the inevitable danger of  loop corrections in these setups involving a sudden and sharp transition from the USR phase to the SR phase. In the right panel of Fig. \ref{totboundary} we have presented $L_c$ as a function of $\Delta N$ for $h=-6$. The value of $L_c$ is calculated from the solution of  $\frac{\Delta \calP_{\mathrm{total}} }{\calP_{\mathrm{cmb}}}|_{L-\mathrm{loop}}=1$  in Eq. (\ref{total-loop}). For example, in this plot, $L_c=1$ and $L_c=2$ corresponds to $\Delta N\simeq2.56$ and $\Delta N\simeq2. 53$ respectively.

\begin{figure}[t]
	\centering
	\includegraphics[ width=0.4\linewidth]{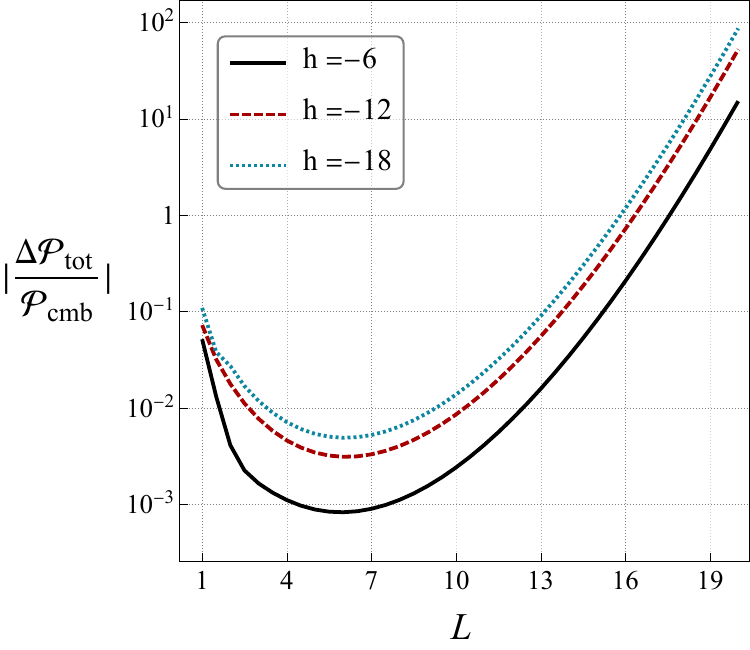}
	\hspace{+1.5 cm} \includegraphics[ width=0.4\linewidth]{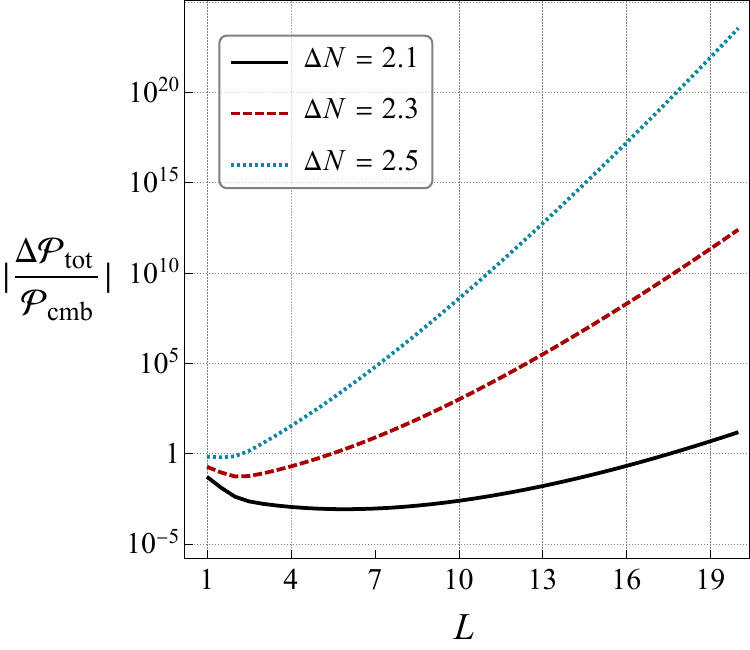}
	\caption{ The total loop corrections as given in  Eq. (\ref{total-loop}).  
	{\bf Left:} Loop corrections with $\Delta N=2.1$ for three values of $h$, from  bottom to top: $h=-6, h=-12$ and $h=-18$. Since $\Delta N< 2.5$, one has to proceed to higher values of $L$ in order for the loop corrections to go out of perturbative control. 
{\bf Right:} The loop corrections for three different values of $\Delta N$, from bottom to top: $\Delta N=2.1, \Delta N=2.3$ and $\Delta N=2.5$ for $h=-6$. This figure  shows the exponential sensitivity of the loop corrections 
to the duration of USR which is encoded in the factor $ \big( \Delta N e^{6 \Delta N} \big)^L$.   }
\label{threeN}
\end{figure}

In Fig. \ref{threeN} we have presented the total loop corrections from Eq. (\ref{total-loop}).  In the left panel of this figure  the total loop correction for $\Delta N=2.1$ is shown for different values of $h$.  As this  value of $\Delta N$ is smaller than $\Delta N\simeq2.5$ required for PBHs formation,  one has to wait longer than 4-loop level for the loop corrections to build up in order to get out of perturbative control.  In the right panel of this figure we have presented the loop corrections for three values of $\Delta N=2.1, 2.3$ and $\Delta N=2.5$ for $h=-6$. We see the strong sensitivity to $\Delta N$ in this plot. This plot also demonstrates the conclusion that the larger is $\Delta N$ the smaller is the critical value $L_c$.

\begin{figure}[t!]
	\centering
	\includegraphics[ width=0.9\linewidth]{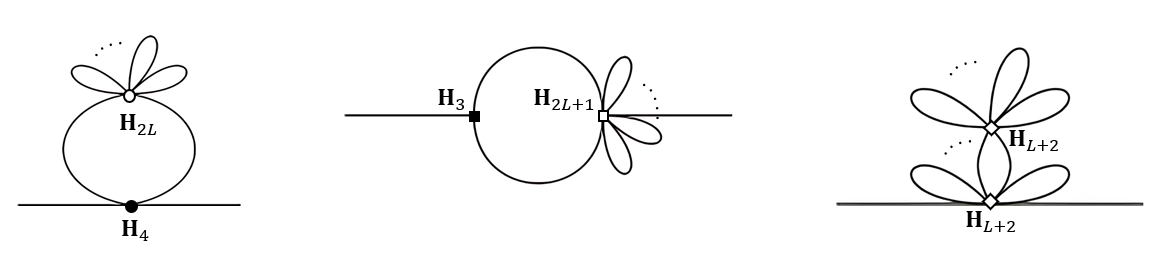}
	\vspace{0.5 cm}
	\caption{ Some  two-vertices Feynman diagrams at $L$-loop order.
	Left: $({\bf H}_{4}, {\bf H}_{2 L})$ vertices, middle: $({\bf H}_{3}, {\bf H}_{2 L+1})$ vertices, right: (${\bf H}_{L+2}, {\bf H}_{L+2})$ vertices.   }
\label{torob}
\end{figure}

So far in our analysis we have considered only one-vertex diagrams as shown in Fig. \ref{diagrams}. For comparison, in Fig. \ref{torob} we have presented some two-vertices diagrams at $L$-loop order.  From the asymptotic $(18 L)^L$ behaviour of  total loop correction Eq. (\ref{approximation}), one concludes that for $L\gg1$ our one-vertex diagrams in Fig. \ref{diagrams} give the dominant loop corrections compared to   multi-vertices $L$-loop diagrams. For example, consider the two-vertices ``torobche"\footnote{Torobche means radish in Persian.}  diagram involving the quartic  ${\bf H}_4$ vertex and the  ${\bf H}_{2 L}$  vertex in which both external  lines are attached to the ${\bf H}_4$ vertex
as depicted in the left panel of Fig. \ref{torob}.  As demonstrated in Appendix \ref{multi}, one can show that the loop correction from 
our single-vertex diagram is larger than that of  the ``torobche" diagram by a factor $18 L$. Similar conclusion can be drawn for other two diagrams in Fig. \ref{torob}, see Appendix \ref{multi} for further details. We comment that  the conclusion that the in-in integrals dominate for one-vertex diagrams compared to multi-vertices diagrams was also noticed in \cite {Leblond:2010yq} who studied the tree-level higher order correlation functions.

The challenging mathematical question is whether the sum of all multi-vertex diagrams, like the one shown in Fig.~5, can combine in such a way that they cancel out the contributions from the one-vertex diagrams. We have no mathematical proof for this question. However, we believe that such a cancellation seems unlikely. The reason is that each multi-vertex diagram is associated with numerical factors \( f(L) (-1)^{g(L)} \), so it is unlikely that they add coherently with each other just to cancel the leading one-vertex diagram. This cancellation requires the presence of a hidden fundamental symmetry, but we do not observe any particular symmetry considerations in the context that would support such a cancellation.
This is a drawback of our perturbative approach based on the Dyson series of the in-in formalism. A mathematically rigorous answer to this question may require non-perturbative analysis based on the path integral approach or Euclidean QFTs. These are interesting but rather complicated questions that are beyond the scope of our current analysis.

\section{Summary and Discussions}
\label{summary}

The setup  of single field USR inflation  has interesting properties such as providing an example for the violation of the non-Gaussianity consistency condition or yielding a mechanism for PBHs formation. 
In this work we have employed the formalism of EFT of inflation to construct the interaction Hamiltonian to an arbitrary order in perturbation theory. In the decoupling limit where the gravitational back-reactions can be neglected, the EFT formalism is very powerful  in calculating the action and the Hamiltonian. 
Neglecting the subleading ${\cal O} (\epsilon^2)$ corrections, we have presented a non-perturbative expression for the interaction Hamiltonian in 
the bulk of USR in terms of the Goldstone field $\pi$ as given in 
Eq. (\ref{HI-eq3}). While the non-perturbative Hamiltonian (\ref{HI-eq3}) is quite interesting on its own right, but  to complete the job, we need a dictionary relating the Goldstone field $\pi$ to the curvature perturbation $\calR$. This non-linear relation between $\pi$ and $\calR$ is provided in  
Eq. (\ref{R-eq2}) to arbitrary orders in perturbation theory. 

As a non-trivial application of our results, we have employed our higher order 
interaction Hamiltonians to calculate the $L$-loop corrections in power spectrum in models of inflation involving an intermediate phases of USR inflation. This analysis extends the previous results obtained for one-loop case 
such as in \cite{Kristiano:2022maq, Firouzjahi:2023aum} and the two-loop analysis partially performed  in  \cite{Firouzjahi:2024sce}.  Among numerous  independent Feynman diagrams at $L$-loop level, we have considered a particular diagram containing a single vertex as shown in Fig. \ref{diagrams}.  
There were two reasons for this consideration. First, since we have a single vertex, then the in-in integrals become somewhat  simplified. 
Otherwise, if one considers a diagram with multiple vertices, then one has to deal with multiple nested time in-in integrals which are far more complicated to handle. Second, as we have argued, this diagram with a single vertex and $L$-loops attached has the dominant contributions in loop corrections. This is controlled roughly by the combination $(18 L)^{ L}$. 

The total loop corrections are the sum of contributions from the bulk of USR as well as from  the point of boundary where the USR phase is glued to the SR phase in an instant and sharp transition. Interestingly, we were able to find a non-perturbative result for the bulk contribution by resumming all loop corrections as given in 
Eq. (\ref{resum-bulk}). On the other hand, the contributions from the 
boundary have the dominant effects.  We have shown that the loop corrections from the boundary terms do not converge 
for $L\gg1$. This is because these contributions are originated from 
the localized sources such as $\dot \eta, \ddot \eta, \dot \eta^2, \dddot \eta, \dot \eta^3$ etc which become increasingly more singular as one considers higher loop levels. The rapid growth of loop corrections induced from the boundary terms is controlled by the parameter $R_{\mathrm{b}}(L) $ which roughly scales as  $(18 L)^L$ for $L\gg1$. 

We have shown that the  $L$-loop corrections scale as 
 $\Big( 18 L  \Delta N e^{6 \Delta N} \calP_{\mathrm{cmb}}\Big)^L$,  indicating strong dependence on the duration of USR phase. It is shown that for a given value of $\Delta N$, there is a critical value $L_c$ above which the loop corrections get out of perturbative control. In the conventional USR setup employed for the PBHs formation  with $\Delta N\sim 2.5$, this happens 
 by $L=4$. This means that this  setup of PBHs formation is not under perturbative control in its simplest realization.         

One may ask why the loop corrections get out of perturbative control as one goes to higher loop levels? We believe the reason is inherited in the 
assumptions of the instant and the sharp transitions from the USR stage to the SR phase. As mentioned above, the sudden jump in $\eta$ induces delta-like singularities which become stronger at higher loop levels. To bypass dangerous loop corrections,  one has to consider a smooth transition from the USR phase to the final attractor phase. In addition, considering an extended relaxation period 
with $|h| \ll 1$ allows the mode function to evolve smoothly so this will erase the large loop corrections in the follow up SR stage. However, these modifications cause the setup to be intractable analytically and a full numerical analysis would be required. It is an interesting question to numerically investigate the higher order loop corrections  in the more physical setup of smooth transition and see whether or not dangerous loop corrections are tamed.  Motivated by the above discussions, we conjecture that the loop corrections in other models of inflation involving potentials with localized features may be important as well. More specifically, as long as the localized feature is 
sharp enough to be approximated by a delta function, then our analysis suggests that the $L$-loop correction is expected to have the form 
$R_f(L) \calP_f^L$ in which $\calP_f$ is the power spectrum at the
point of localized feature and $R_f(L)$ is a parameter like our $R_b(L)$. 
It would be interesting to examine this conjecture in models involving sharp features such as in models with rapid turn in field space. 

We have not studied the renormalization of the loop corrections. Indeed, renormalization is an important step to obtain the finite physical quantities.
We believe that our main conclusions summarized above, such as  the order of the loop corrections and the sensitivity of the loop corrections  to $\Delta N$ etc, remain intact even including the effects of renormalization. The reason is that there is no underlying symmetry to cancel all loop corrections at each order of $L$. The renormalization of one-loop corrections was studied in \cite{Sheikhahmadi:2024peu} with the conclusion that the loop corrections remain still dangerous  after the renormalization is taken into account. More specifically, the amplitude of the renormalized power spectrum is still given by the quantity  $\calP_e$, i.e the peak of power at the end of USR, so the loop corrections will get out of perturbative control when the transition is sharp and $\Delta N$ is large enough. 

\vspace{0.7cm}

{\bf Note added:} 
After the completion of our work here and in the more recent companion paper \cite{Firouzjahi:2025gja}, we became aware of the works \cite{Behbahani:2011it} and \cite{Creminelli:2024cge} \footnote{ We thank A. A. Abolhasani for bringing Ref.  \cite{Behbahani:2011it} to our attention when our work was complete. We thank S. Renaux-Petel for bringing Ref. \cite{Creminelli:2024cge} to our attention after our companion paper \cite{Firouzjahi:2025gja} appeared in arXiv.}. In \cite{Behbahani:2011it},  the authors 
calculated $S_{\pi^n}$ similar to our Eq. (\ref{eq:Sn final0}), 
in the setup of inflation with discrete shift symmetry. 
In \cite{Creminelli:2024cge}, the authors presented the non-perturbative action Eq. (\ref{action-pi}) in the general single field model in the decoupling limit.  It is also concluded in  \cite{Creminelli:2024cge} that $\pi= \mathrm{constant}$ is a solution to all order. This is true in our USR setup as well, but the important point is that neither $\pi$ nor $\calR$ are conserved on superhorizon scales. As elaborated around 
our Eq. (\ref{R-sol}), the would-be decaying mode is actually the dominant solution with $\calR \propto \pi  \propto a^{3}$.

\vspace{0.7cm}

{\bf Acknowledgments:}  We thank A. A.  Abolhasani,  M. H. Namjoo, S. Renaux-Petel,  A. Riotto and    H. Sheikhahmadi for useful discussions and correspondences.  The work of H. F. is supported  by INSF of Iran under the grant number 4038049. B. N. thanks ICTP for hospitality when this work was in progress.

\vspace{0.7cm}

\appendix

\section{Mode Functions}
\label{mod-functions}

In this Appendix, we present the mode functions   which are required to perform the in-in integrals.  

As usual we decompose the mode function of the curvature perturbation in Fourier space in terms of the creation and annihilation operator as follows, 
\begin{equation}
\calR({\bf x}, t) = \int \frac{d^3 k}{(2\pi)^3} e^{i {\bf k}\cdot {\bf x}} \hat\calR_{\bf k}(t) \, ,
\end{equation}
in which  $\hat\calR_{\bf k}(t)= \calR_k(t) a_{\bf k} + \calR^*_k(t) a_{-\bf k}^\dagger$ and    
$[ a_{\bf k}, a^\dagger_{\bf k'} ] = ( 2 \pi)^3 \delta (  {\bf k} - {\bf k'}) $.

Starting with an initial Bunch-Davies (Minkowski) in the first SR phase, 
\begin{equation}
\calR^{(1)}_{k} =  \frac{H}{ M_P\sqrt{4 \epsilon_i k^3}}
( 1+ i k \tau) e^{- i k \tau} \, , \quad \quad (\tau < \tau_s) \, ,
\end{equation}
the mode function in the USR phase is obtained by imposing the continuity of 
$\calR$ and its derivative at $\tau=\tau_s$, yielding 
\begin{equation}
\calR^{(2)}_{k} =  \frac{H}{ M_P\sqrt{4 \epsilon_i k^3}}  \bigg( \frac{\tau_s}{\tau} \bigg)^3
\Big[ \alpha^{(2)}_k ( 1+ i k \tau) e^{- i k \tau}  + \beta^{(2)}_k ( 1- i k \tau) e^{ i k \tau}  \Big]  \, ,
\end{equation}
with the  coefficients $\alpha^{(2)}_k$ and $\beta^{(2)}_k$ given by,
\begin{equation}
\label{alpha-beta2}
\alpha^{(2)}_k = 1 + \frac{3 i }{ 2 k^3 \tau_s^3} ( 1 + k^2 \tau_s^2) \, , \quad \quad
\beta^{(2)}_k= -\frac{3i }{ 2 k^3 \tau_s^3 } {( 1+ i k \tau_s)^2} e^{- 2 i k \tau_s} \, .
\end{equation}
Imposing the matching condition at $\tau=\tau_e$, the outgoing 
mode function in the third phase is obtained to be, 
\begin{equation}
\calR^{(3)}_{k} =  \frac{H}{ M_P\sqrt{4 \epsilon(\tau) k^3}}
\Big[ \alpha^{(3)}_k ( 1+ i k \tau) e^{- i k \tau}  + \beta^{(3)}_k ( 1- i k \tau) e^{ i k \tau}  \Big] \, ,
\end{equation}
in which the coefficients  $\alpha^{(3)}_k$ and $\beta^{(3)}_k$ are given by,
\begin{equation}
\label{alpha3-beta3}
\begin{pmatrix}
\alpha^{(3)}_k \\
\\
\beta^{(3)}_k
\end{pmatrix}
=
\frac{1}{4 k^3 \tau_e^3}
\begin{pmatrix}
\gamma_{11} \quad \gamma_{12}\\
\\
\gamma_{21} \quad \gamma_{22}
\end{pmatrix}
\begin{pmatrix}
\alpha^{(2)}_k \\
\\
\beta^{(2)}_k
\end{pmatrix},
\end{equation}
 where,
\begin{equation}
\begin{gathered}
\label{eq:gammas}
\gamma_{11} = \gamma^*_{22} = 4 k^3 \tau_e^3 + i h \left( 1+  k^2 \tau_e^2 \right),\\
\\
\gamma_{12} = \gamma^*_{21} = i h e^{2 i k \tau_e} \left(1 -i k \tau_e \right)^2. 
\end{gathered}
\end{equation}

In particular, note the contributions of the sharpness parameter $h$ which appears in various places in coefficients  $\alpha^{(3)}_k$ and $\beta^{(3)}_k$. The above mode functions are used to perform the in-in integrals for the loop corrections.


\section{Relation Between $\tilde \epsilon^{(2 L)}$ and 
$\epsilon^{(2 L)}$}
\label{eps-relation}

In the main text we have  calculated the contribution of the gradient boundary term $\tilde A_1$ involving $\epsilon^{(2 L)}(t)$. Correspondingly, one hopes to perform similar steps for the kinetic terms $A_1, A_2$ and $B$ which involve $\tilde \epsilon^{(2 L)}(t)$. However, it turns out that the analysis for $\tilde \epsilon^{(2 L)}(t)$ are more difficult to handle. 
To see the difficulties associated with $\tilde \epsilon^{(2 L)}(t)$, let us look at its expression,
\ba
\label{tep-N2}
\tilde{\epsilon}^{(2 L)}(t)  = H^{2 L}  
\Big(  \epsilon_1(N)\Theta(-N) + \epsilon_2(N) \Theta(N) \Big)^2 
\Big(  \epsilon_1^{-1}(N) \Theta(-N) + \epsilon_2^{-1}(N) \Theta(N)  \Big)^{(2L)}  + {\cal O} (\epsilon^2) \,
\ea
The trouble with this expression is that we can not simply perform the integrations by parts as in the case of of $\tilde A_1$ described after 
Eq. (\ref{tildeA1-int}). This is because after taking one integration by parts, we transfer the derivative over the left term involving $\Theta(N)$ and $\Theta(-N)$ which brings back the delta functions and its derivatives.

To bypass this difficulties, let us look at the analytic realization of the $\Theta$ function as follows,
\ba 
\label{real1}
\Theta(N) = \lim_ {~~ k \rightarrow \infty}  \frac{1}{1+ \exp(-2 k N)} \, .
\ea
With the above prescription, we obtain,
\ba
\epsilon(N)= \epsilon_e\, \lim_ {  k \rightarrow \infty} \frac{ e^{ (2k+ \eta_2)N} + e^{\eta_1 N} }{1+ e^{ 2 k N} } \, .
\ea
Constructing the quantities $\epsilon^{(m)}(N)$ and 
$\tilde \epsilon^{(m)}(N)$ from the above expressions, one finds that,
\ba
\big( \epsilon^{(m)}+ (-1)^m  \tilde \epsilon^{(m)} \big)  \Big|_{N=0}= 
\epsilon_e \big( \eta_1^{m}+ \eta_2^{m} \big) \, .
\ea 
The key point to observe is that the right hand side of the above equation 
is independent of $k$ and 
is finite so its effect  in an integration across the boundary is zero. 
Therefore, as far as the boundary terms are concerned,  one can use the following prescription,
\ba
\label{prescription0}
\epsilon^{(m)}  \cong  (-1)^{m-1} \tilde \epsilon^{(m)} \, ,  \quad \quad (N=0) \, .
\ea
In particular, for our case with $m=2L$, we obtain 
\ba
\label{prescription1}
\epsilon^{(2 L)}  \cong  - \tilde \epsilon^{(2 L)} \, ,  \quad \quad (N=0) \, .
\ea
Here the symbol $\cong$ indicates that this is not an identity but a 
prescription as far as the singular terms are concerned. With the above relation between $\tilde \epsilon^{(2 L)}$ and  $\epsilon^{(2 L)}$, one can calculate the kinetic terms $A_1, A_2$ and $B$ similar to the case of 
$\tilde A_1$. 

One may wonder if the  conclusion (\ref{prescription1}) is limited to the particular realization of the step function given in Eq. (\ref{real1}). We have checked that our conclusion is independent of this particular realization.
For example, the following two realizations of the step function, 
\ba
\label{real2}
\Theta(N) = \lim_ {~~ k \rightarrow \infty}  \Big( \frac{1}{2} + \frac{1}{\pi}
\mathrm{arctan} (k N) \Big) \, ,
\ea
and 
\ba
\label{real3}
\Theta(N) = \lim_ {~~ k \rightarrow \infty}  \Big( \frac{1}{2} + \frac{1}{2}
\mathrm{erf} (k N) \Big)  \, ,
\ea
yield  to the same conclusion as Eq. (\ref{prescription1}). 

This can be stemmed from the mathematical property that the $\Theta$ function can be viewed as a constant plus an odd function. Consequently, its odd (even) derivatives produce even (odd) functions. When integrating over the transition, we can disregard the odd functions. Therefore, we only need to consider the even functions in both $\epsilon^{(m)}$ and $\tilde{\epsilon}^{(m)}$, and the even functions are related via Eq.~\eqref{prescription0}.


\section{Summation for Boundary Terms at Order $L$}
\label{sum-L}

In this appendix, we present the sum over $k$ for the boundary terms 
$A_1, A_2, B$ and $\tilde A_1$ as given in Eqs. (\ref{A1-te})-(\ref{tA1-te}). 
 
Let us start with $A_1$ term. Defining the fractional loop corrections induced by this term via
\ba
\frac{\Delta \calP^{(A_1)}_{\mathrm{boundary}} }{\calP_{\mathrm{cmb}}}\Big|_{L-\mathrm{loop}} \equiv R_{\mathrm{b}}^{(A_1)} (L) 
\big( \Delta N e^{6 \Delta N}  \calP_{\mathrm{cmb}}\big)^L \, ,
\ea
we obtain, 
\ba
\label{RA1}
R_{\mathrm{b}}^{(A_1)} (L) &= &\frac{6 }{2^L \Gamma(L) }
\sum_{k=1}^{2 L } \Big(  (6+h)^{2L-k} - 6^{2L-k}
\Big) \Big[   (6+ 6 L)^{k-1}+ \frac{(6-h)}{h} ( 6L+ 3)^{k-1} \Big]  \nonumber\\
  & = & \frac{2^{-L}}{h L \left( h - 6 L\right) \left( 3 + h - 6 L\right) \left( 2 L -1 \right) \Gamma (L) } \Big\{ 54\, L (6 + h)^{2 L} (h - 4 L) (2 L -1)  \nonumber\\
  & +&  36^L (6 L -h-3) \Big[ h^2 (1+L)^{2 L} (2 L -1) +  (12 L-h)(6 L-h) \Big]   \nonumber \\
  &+&  2 \times 9^L h L (h-6) (6 L -h) (1 +2 L)^{2 L}  \Big\} \, .
\ea
The contribution of $A_2$ is simply related to $A_1$ via, 
\ba
\label{RA2}
R_{\mathrm{b}}^{(A_2)} (L) &=& 2 (L-1) R_{\mathrm{b}}^{(A_1)} (L) .
\ea

On the other hands, for the other terms, we obtain
\ba
\label{RB}
R_{\mathrm{b}}^{(B)} (L) &=& 
 \frac{12}{2^L \Gamma(L) }
\sum_{k=1}^{2 L } \Big[  (6+h)^{2L-k} - 6^{2L-k} \Big]   (6+ 6 L)^{k-1} 
  \nonumber\\
&=& \frac{2^{1-L} \Big\{36^L \Big[h \left((L+1)^{2 L}-1\right)+6 L\Big]-6 L (h+6)^{2 L}\Big\}}{L (6 L-h) \Gamma (L)} \, ,
\ea
and
\ba
\label{tA1}
R_{\mathrm{b}}^{(\tilde A_1)} (L) &=& 
 \frac{-1}{3\Delta N  2^L \Gamma(L) }
\sum_{k=1}^{2 L } \Big(  (6+h)^{2L-k} - 6^{2L-k} \Big)   
\Big[  (4+ 6 L)^{k-1}+ \frac{(6-h)}{h} ( 6L+ 1)^{k-1} \Big] 
  \nonumber\\
&=& \frac{2^{-L-1}}{3 \Delta N h (3 L-1) (6 L-5) (h-6 L+2) (h-6 L+5) \Gamma (L)} \Big\{h^3 \Big[-6 L (6 L+1)^{2 L}  \nonumber\\
&+&  2 (6 L+1)^{2 L}+3\times 36^L+(6 L-5) (6 L+4)^{2 L}\Big]+h^2 \Big[36 L^2 (6 L+1)^{2 L} \nonumber \\ &+&12 L (6 L+1)^{2 L} 
-  8 (6 L+1)^{2 L}+33\times 36^L-(5-6 L)^2 (6 L+4)^{2 L}-72\times 36^L L\Big]  \nonumber\\
&+&  6 h (3 L-1) \Big[-3 (6 L-5) (h+6)^{2 L}-19\times 36^L-12 L (6 L+1)^{2 L}+4 (6 L+1)^{2 L}\nonumber\\
&+&30\times 36^L L\Big] -  24 (6 L-5) (1-3 L)^2 \Big[36^L-(h+6)^{2 L}\Big]
\Big\} .
\ea

Combining the above four terms, we obtain
\ba
R_{\mathrm{b}} (L)&\equiv & R_{\mathrm{b}}^{(A_1)} (L) + R_{\mathrm{b}}^{(A_2)} (L) + R_{\mathrm{b}}^{(B)} (L)+ R_{\mathrm{b}}^{(\tilde A_1)} (L) \, ,
\ea 
which is used for the total boundary contribution given in Eq. (\ref{boundary-cont}).

Looking at the large $L$ limit of the above results, one can check that the leading contribution comes from $A_2$ term with the relation (\ref{RA2}). For 
$L \gg 1$, one obtains 
\ba
R_{\mathrm{b}}(L)  \rightarrow \frac{\sqrt{2}(6+ (e-1)h) }{6 \sqrt{\pi}}
\Big ( 18^L e^{L+1} L^{L-\frac{1}{2}} \Big)  \sim ( 18 L)^L \,  
\quad \quad (L\gg 1) \, .
\ea

\section{Multi-vertices Diagrams}

\label{multi}

In our analysis, we have considered the one-vertex diagrams 
for $L$-loop corrections as shown in Fig. \ref{diagrams}. As discussed previously, the main reason was that the in-in analysis for these diagrams are easier to handle as they involve a single time in-in integral. In addition, it can be checked that the loop corrections from these diagrams are larger than all other multi-vertices $L$-loop diagrams for  $L\gg1$. This can be seen from the asymptotic $(18L)^L$ behaviour of  total loop correction Eq. (\ref{approximation}). To see this explicitly, consider the two-vertices ``torobche" diagram involving  ${\bf H}_4$ and  ${\bf H}_{2 L}$  vertices where both external  lines are attached to the ${\bf H}_4$ vertex. This diagram is plotted in left panel of Fig. \ref{torob}.

Let us consider the the kinetic terms of the Hamiltonians. Taking into account the numerical prefactors and the contributions of $\tilde\epsilon^{(2 L-2)}$ and 
$\tilde\epsilon^{(2)}$ terms associated to each vertex, the contraction of the quantum operators schematically has the following structure, 
\ba
\frac{\tilde\epsilon^{(2 L-2)}(\tau_1)}{(2 L-2)!} \frac{\tilde\epsilon^{(2)}(\tau_2)}{2!}
  \Big \langle 
 \hat\pi_{\bfk_1} \hat\pi_{\bfk_2}... \hat\pi_{\bfk_{2L-2}}(\tau_1)\, \,  \hat \pi'_{\bfk_{2L-1} } \hat \pi'_{\bfk_{2L}}(\tau_1)  \, \,   
 \pi_{\bfm_1} \hat \pi_{\bfm_2}(\tau_2)  \pi'_{\bfm_3} \hat \pi'_{\bfm_4}(\tau_2) 
 \, \, \hat\pi_{\bf p_1} \hat\pi_{\bf p_2}(\tau_0)
 \Big \rangle   \, .
 \ea
Note that the operators at $\tau_1$ are from ${\bf H}_{2 L}$  while those at 
$\tau_2$ are from ${\bf H}_4$. 
We are interested only in the boundary terms where $\tau_1= \tau_2= \tau_e$.

The dependence on $\Delta N\calP_e$ would be the same as in the case of 
single vertex, yielding to the expected contribution $\big( \Delta N\calP_e \big)^L$. The difference compared to our single vertex case will be in the symmetry factor and the total number of Wick contractions. 

Now consider the term like $A_1$ in our one-vertex diagram in which 
$\pi'_{\bfk_i}$ contract with each other. The number of independent contractions is,
\ba
2\times  (2L-2) (2 L-3)  (2 L-5)!!= 2 (2 L-2) (2 L-3)!! \, .
\ea
Correspondingly, the numerical factors involved are,
\ba
\frac{1}{2!} \frac{1}{(2L-2)!}\times 2 (2 L-2) (2 L-3)!! = \frac{1}{( 2 L-4)!!} \, .
\ea
On the other hand, from the integrations by parts to get rid of 
$\tilde\epsilon^{(2 L-2)}$, the corresponding sum over $k$ in Eq. (\ref{RA1}) 
now runs from $1$ to $ 2L-2$. Considering all these steps, the result now will be similar to the case $A_1$ in Eq. (\ref{RA1}) except replacing $R^{(A_1)}(L)$ by $R^{(A_1)}(L-1)$. Considering other terms like $A_2, B$ etc we see the same relation between the numerical factors.

From the above analysis  we conclude that the loop correction from the ``torobche" diagram is roughly given by 
$R_b(L-1) (\Delta N \calP_e)^L$. Using the asymptotic behaviour 
$R_b(L) \sim (1 8 L)^L$, we conclude that the loop correction from 
``torobche" diagram is smaller than that of our one-vertex diagram by a factor $1/18 L$.

As another example of a two-vertices diagram for $L$-loop corrections, consider the middle 
diagram in Fig. \ref{torob} in which the left vertex is from ${\bf H}_3$ and the right vertex is from ${\bf H}_{2 L+1}$. The number of equivalent contractions is,
\ba
(3) (2 L-1) ( 2L-2) ( 2 L-3) ( 2L-5)!!= 3\times (2L-2) ( 2L-1)!! \, .
\ea
Considering the numerical prefactor $1/(2 L-1)!$ from the Hamiltonian ${\bf H}_{2 L+1}$, the numerical factor involved is given by,
\ba
\frac{3}{( 2 L-1)!} (2L-2) ( 2L-1)!! \, .
\ea
On the other hand, the process of integrations by parts to get rid of $\tilde\epsilon^{(2 L-1)}$ gives a sum  of the form $\sum_{k=1}^{2 L-1}$. Combining all effects into account, and using the asymptotic form of $R_b(L)$, one concludes that the loop correction from this diagram is smaller than our one-vertex diagram roughly by a factor $L^{-3/2}$. 

As the last example of multi-vertices $L$-loop corrections, consider the third diagram in  Fig. \ref{torob} involving two vertices of Hamiltonian ${\bf H}_{L+2}$. Counting the number of Wick contractions and considering the numerical prefactors, one can show that the loop correction from this diagram is smaller than that of our one-vertex diagram roughly by a factor $2^{-L}$. 


\bibliography{Loops}{}

\providecommand{\href}[2]{#2}\begingroup\raggedright\begin{thebibliography}{10}

\bibitem{Maldacena:2002vr}
J.~M. Maldacena, \emph{{Non-Gaussian features of primordial fluctuations in
  single field inflationary models}},
  \href{https://doi.org/10.1088/1126-6708/2003/05/013}{\emph{JHEP} {\bfseries
  05} (2003) 013}, [\href{https://arxiv.org/abs/astro-ph/0210603}{{\ttfamily
  astro-ph/0210603}}].

\bibitem{Jarnhus:2007ia}
P.~R. Jarnhus and M.~S. Sloth, \emph{{de Sitter limit of inflation and
  nonlinear perturbation theory}},
  \href{https://doi.org/10.1088/1475-7516/2008/02/013}{\emph{JCAP} {\bfseries
  02} (2008) 013}, [\href{https://arxiv.org/abs/0709.2708}{{\ttfamily
  0709.2708}}].

\bibitem{Arroja:2008ga}
F.~Arroja and K.~Koyama, \emph{{Non-gaussianity from the trispectrum in general
  single field inflation}},
  \href{https://doi.org/10.1103/PhysRevD.77.083517}{\emph{Phys. Rev. D}
  {\bfseries 77} (2008) 083517},
  [\href{https://arxiv.org/abs/0802.1167}{{\ttfamily 0802.1167}}].

\bibitem{Cheung:2007st}
C.~Cheung, P.~Creminelli, A.~L. Fitzpatrick, J.~Kaplan and L.~Senatore,
  \emph{{The Effective Field Theory of Inflation}},
  \href{https://doi.org/10.1088/1126-6708/2008/03/014}{\emph{JHEP} {\bfseries
  03} (2008) 014}, [\href{https://arxiv.org/abs/0709.0293}{{\ttfamily
  0709.0293}}].

\bibitem{Cheung:2007sv}
C.~Cheung, A.~L. Fitzpatrick, J.~Kaplan and L.~Senatore, \emph{{On the
  consistency relation of the 3-point function in single field inflation}},
  \href{https://doi.org/10.1088/1475-7516/2008/02/021}{\emph{JCAP} {\bfseries
  02} (2008) 021}, [\href{https://arxiv.org/abs/0709.0295}{{\ttfamily
  0709.0295}}].

\bibitem{Kristiano:2022maq}
J.~Kristiano and J.~Yokoyama, \emph{{Constraining Primordial Black Hole
  Formation from Single-Field Inflation}},
  \href{https://doi.org/10.1103/PhysRevLett.132.221003}{\emph{Phys. Rev. Lett.}
  {\bfseries 132} (2024) 221003},
  [\href{https://arxiv.org/abs/2211.03395}{{\ttfamily 2211.03395}}].

\bibitem{Kristiano:2023scm}
J.~Kristiano and J.~Yokoyama, \emph{{Note on the bispectrum and one-loop
  corrections in single-field inflation with primordial black hole formation}},
  \href{https://doi.org/10.1103/PhysRevD.109.103541}{\emph{Phys. Rev. D}
  {\bfseries 109} (2024) 103541},
  [\href{https://arxiv.org/abs/2303.00341}{{\ttfamily 2303.00341}}].

\bibitem{Riotto:2023hoz}
A.~Riotto, \emph{{The Primordial Black Hole Formation from Single-Field
  Inflation is Not Ruled Out}},
  \href{https://arxiv.org/abs/2301.00599}{{\ttfamily 2301.00599}}.

\bibitem{Riotto:2023gpm}
A.~Riotto, \emph{{The Primordial Black Hole Formation from Single-Field
  Inflation is Still Not Ruled Out}},
  \href{https://arxiv.org/abs/2303.01727}{{\ttfamily 2303.01727}}.

\bibitem{Choudhury:2023vuj}
S.~Choudhury, M.~R. Gangopadhyay and M.~Sami, \emph{{No-go for the formation of
  heavy mass Primordial Black Holes in Single Field Inflation}},
  \href{https://doi.org/10.1140/epjc/s10052-024-13218-2}{\emph{Eur. Phys. J. C}
  {\bfseries 84} (2024) 884},
  [\href{https://arxiv.org/abs/2301.10000}{{\ttfamily 2301.10000}}].

\bibitem{Choudhury:2023jlt}
S.~Choudhury, S.~Panda and M.~Sami, \emph{{PBH formation in EFT of single field
  inflation with sharp transition}},
  \href{https://doi.org/10.1016/j.physletb.2023.138123}{\emph{Phys. Lett. B}
  {\bfseries 845} (2023) 138123},
  [\href{https://arxiv.org/abs/2302.05655}{{\ttfamily 2302.05655}}].

\bibitem{Choudhury:2023rks}
S.~Choudhury, S.~Panda and M.~Sami, \emph{{Quantum loop effects on the power
  spectrum and constraints on primordial black holes}},
  \href{https://doi.org/10.1088/1475-7516/2023/11/066}{\emph{JCAP} {\bfseries
  11} (2023) 066}, [\href{https://arxiv.org/abs/2303.06066}{{\ttfamily
  2303.06066}}].

\bibitem{Choudhury:2023hvf}
S.~Choudhury, S.~Panda and M.~Sami, \emph{{Galileon inflation evades the no-go
  for PBH formation in the single-field framework}},
  \href{https://doi.org/10.1088/1475-7516/2023/08/078}{\emph{JCAP} {\bfseries
  08} (2023) 078}, [\href{https://arxiv.org/abs/2304.04065}{{\ttfamily
  2304.04065}}].

\bibitem{Choudhury:2024one}
S.~Choudhury, A.~Karde, S.~Panda and M.~Sami, \emph{{Realisation of the
  ultra-slow roll phase in Galileon inflation and PBH overproduction}},
  \href{https://doi.org/10.1088/1475-7516/2024/07/034}{\emph{JCAP} {\bfseries
  07} (2024) 034}, [\href{https://arxiv.org/abs/2401.10925}{{\ttfamily
  2401.10925}}].

\bibitem{Choudhury:2024aji}
S.~Choudhury and M.~Sami, \emph{{Large fluctuations and Primordial Black
  Holes}}, \href{https://doi.org/10.1016/j.physrep.2024.10.007}{\emph{Phys.
  Rept.} {\bfseries 1103} (2025) 1--276},
  [\href{https://arxiv.org/abs/2407.17006}{{\ttfamily 2407.17006}}].

\bibitem{Firouzjahi:2023aum}
H.~Firouzjahi, \emph{{One-loop corrections in power spectrum in single field
  inflation}}, \href{https://doi.org/10.1088/1475-7516/2023/10/006}{\emph{JCAP}
  {\bfseries 10} (2023) 006},
  [\href{https://arxiv.org/abs/2303.12025}{{\ttfamily 2303.12025}}].

\bibitem{Motohashi:2023syh}
H.~Motohashi and Y.~Tada, \emph{{Squeezed bispectrum and one-loop corrections
  in transient constant-roll inflation}},
  \href{https://doi.org/10.1088/1475-7516/2023/08/069}{\emph{JCAP} {\bfseries
  08} (2023) 069}, [\href{https://arxiv.org/abs/2303.16035}{{\ttfamily
  2303.16035}}].

\bibitem{Firouzjahi:2023ahg}
H.~Firouzjahi and A.~Riotto, \emph{{Primordial Black Holes and loops in
  single-field inflation}},
  \href{https://doi.org/10.1088/1475-7516/2024/02/021}{\emph{JCAP} {\bfseries
  02} (2024) 021}, [\href{https://arxiv.org/abs/2304.07801}{{\ttfamily
  2304.07801}}].

\bibitem{Tasinato:2023ukp}
G.~Tasinato, \emph{{Large |\ensuremath{\eta}| approach to single field
  inflation}}, \href{https://doi.org/10.1103/PhysRevD.108.043526}{\emph{Phys.
  Rev. D} {\bfseries 108} (2023) 043526},
  [\href{https://arxiv.org/abs/2305.11568}{{\ttfamily 2305.11568}}].

\bibitem{Franciolini:2023agm}
G.~Franciolini, A.~Iovino, Junior., M.~Taoso and A.~Urbano,
  \emph{{Perturbativity in the presence of ultraslow-roll dynamics}},
  \href{https://doi.org/10.1103/PhysRevD.109.123550}{\emph{Phys. Rev. D}
  {\bfseries 109} (2024) 123550},
  [\href{https://arxiv.org/abs/2305.03491}{{\ttfamily 2305.03491}}].

\bibitem{Firouzjahi:2023btw}
H.~Firouzjahi, \emph{{Loop corrections in gravitational wave spectrum in single
  field inflation}},
  \href{https://doi.org/10.1103/PhysRevD.108.043532}{\emph{Phys. Rev. D}
  {\bfseries 108} (2023) 043532},
  [\href{https://arxiv.org/abs/2305.01527}{{\ttfamily 2305.01527}}].

\bibitem{Maity:2023qzw}
S.~Maity, H.~V. Ragavendra, S.~K. Sethi and L.~Sriramkumar, \emph{{Loop
  contributions to the scalar power spectrum due to quartic order action in
  ultra slow roll inflation}},
  \href{https://doi.org/10.1088/1475-7516/2024/05/046}{\emph{JCAP} {\bfseries
  05} (2024) 046}, [\href{https://arxiv.org/abs/2307.13636}{{\ttfamily
  2307.13636}}].

\bibitem{Cheng:2023ikq}
S.-L. Cheng, D.-S. Lee and K.-W. Ng, \emph{{Primordial perturbations from
  ultra-slow-roll single-field inflation with quantum loop effects}},
  \href{https://doi.org/10.1088/1475-7516/2024/03/008}{\emph{JCAP} {\bfseries
  03} (2024) 008}, [\href{https://arxiv.org/abs/2305.16810}{{\ttfamily
  2305.16810}}].

\bibitem{Fumagalli:2023loc}
J.~Fumagalli, S.~Bhattacharya, M.~Peloso, S.~Renaux-Petel and L.~T. Witkowski,
  \emph{{One-loop infrared rescattering by enhanced scalar fluctuations during
  inflation}}, \href{https://doi.org/10.1088/1475-7516/2024/04/029}{\emph{JCAP}
  {\bfseries 04} (2024) 029},
  [\href{https://arxiv.org/abs/2307.08358}{{\ttfamily 2307.08358}}].

\bibitem{Nassiri-Rad:2023asg}
A.~Nassiri-Rad and K.~Asadi, \emph{{Induced gravitational waves from
  non-attractor inflation and NANOGrav data}},
  \href{https://doi.org/10.1088/1475-7516/2024/04/009}{\emph{JCAP} {\bfseries
  04} (2024) 009}, [\href{https://arxiv.org/abs/2310.11427}{{\ttfamily
  2310.11427}}].

\bibitem{Meng:2022ixx}
D.-S. Meng, C.~Yuan and Q.-g. Huang, \emph{{One-loop correction to the enhanced
  curvature perturbation with local-type non-Gaussianity for the formation of
  primordial black holes}},
  \href{https://doi.org/10.1103/PhysRevD.106.063508}{\emph{Phys. Rev. D}
  {\bfseries 106} (2022) 063508},
  [\href{https://arxiv.org/abs/2207.07668}{{\ttfamily 2207.07668}}].

\bibitem{Cheng:2021lif}
S.-L. Cheng, D.-S. Lee and K.-W. Ng, \emph{{Power spectrum of primordial
  perturbations during ultra-slow-roll inflation with back reaction effects}},
  \href{https://doi.org/10.1016/j.physletb.2022.136956}{\emph{Phys. Lett. B}
  {\bfseries 827} (2022) 136956},
  [\href{https://arxiv.org/abs/2106.09275}{{\ttfamily 2106.09275}}].

\bibitem{Fumagalli:2023hpa}
J.~Fumagalli, \emph{{Absence of one-loop effects on large scales from small
  scales in non-slow-roll dynamics}},
  \href{https://arxiv.org/abs/2305.19263}{{\ttfamily 2305.19263}}.

\bibitem{Tada:2023rgp}
Y.~Tada, T.~Terada and J.~Tokuda, \emph{{Cancellation of quantum corrections on
  the soft curvature perturbations}},
  \href{https://doi.org/10.1007/JHEP01(2024)105}{\emph{JHEP} {\bfseries 01}
  (2024) 105}, [\href{https://arxiv.org/abs/2308.04732}{{\ttfamily
  2308.04732}}].

\bibitem{Firouzjahi:2023bkt}
H.~Firouzjahi, \emph{{Revisiting loop corrections in single field
  ultraslow-roll inflation}},
  \href{https://doi.org/10.1103/PhysRevD.109.043514}{\emph{Phys. Rev. D}
  {\bfseries 109} (2024) 043514},
  [\href{https://arxiv.org/abs/2311.04080}{{\ttfamily 2311.04080}}].

\bibitem{Iacconi:2023slv}
L.~Iacconi and D.~J. Mulryne, \emph{{Multi-field inflation with large scalar
  fluctuations: non-Gaussianity and perturbativity}},
  \href{https://doi.org/10.1088/1475-7516/2023/09/033}{\emph{JCAP} {\bfseries
  09} (2023) 033}, [\href{https://arxiv.org/abs/2304.14260}{{\ttfamily
  2304.14260}}].

\bibitem{Davies:2023hhn}
M.~W. Davies, L.~Iacconi and D.~J. Mulryne, \emph{{Numerical 1-loop correction
  from a potential yielding ultra-slow-roll dynamics}},
  \href{https://doi.org/10.1088/1475-7516/2024/04/050}{\emph{JCAP} {\bfseries
  04} (2024) 050}, [\href{https://arxiv.org/abs/2312.05694}{{\ttfamily
  2312.05694}}].

\bibitem{Iacconi:2023ggt}
L.~Iacconi, D.~Mulryne and D.~Seery, \emph{{Loop corrections in the separate
  universe picture}},
  \href{https://doi.org/10.1088/1475-7516/2024/06/062}{\emph{JCAP} {\bfseries
  06} (2024) 062}, [\href{https://arxiv.org/abs/2312.12424}{{\ttfamily
  2312.12424}}].

\bibitem{Kawaguchi:2024lsw}
R.~Kawaguchi, S.~Tsujikawa and Y.~Yamada, \emph{{Roles of boundary and
  equation-of-motion terms in cosmological correlation functions}},
  \href{https://doi.org/10.1016/j.physletb.2024.138962}{\emph{Phys. Lett. B}
  {\bfseries 856} (2024) 138962},
  [\href{https://arxiv.org/abs/2403.16022}{{\ttfamily 2403.16022}}].

\bibitem{Braglia:2024zsl}
M.~Braglia and L.~Pinol, \emph{{No time to derive: unraveling total time
  derivatives in in-in perturbation theory}},
  \href{https://doi.org/10.1007/JHEP08(2024)068}{\emph{JHEP} {\bfseries 08}
  (2024) 068}, [\href{https://arxiv.org/abs/2403.14558}{{\ttfamily
  2403.14558}}].

\bibitem{Firouzjahi:2024psd}
H.~Firouzjahi, \emph{{Loop corrections in the bispectrum in ultraslow-roll
  inflation with PBHs formation}},
  \href{https://doi.org/10.1103/PhysRevD.110.043519}{\emph{Phys. Rev. D}
  {\bfseries 110} (2024) 043519},
  [\href{https://arxiv.org/abs/2403.03841}{{\ttfamily 2403.03841}}].

\bibitem{Caravano:2024moy}
A.~Caravano, G.~Franciolini and S.~Renaux-Petel, \emph{{Ultra-Slow-Roll
  Inflation on the Lattice: Backreaction and Nonlinear Effects}},
  \href{https://arxiv.org/abs/2410.23942}{{\ttfamily 2410.23942}}.

\bibitem{Caravano:2024tlp}
A.~Caravano, K.~Inomata and S.~Renaux-Petel, \emph{{Inflationary Butterfly
  Effect: Nonperturbative Dynamics from Small-Scale Features}},
  \href{https://doi.org/10.1103/PhysRevLett.133.151001}{\emph{Phys. Rev. Lett.}
  {\bfseries 133} (2024) 151001},
  [\href{https://arxiv.org/abs/2403.12811}{{\ttfamily 2403.12811}}].

\bibitem{Saburov:2024und}
S.~Saburov and S.~V. Ketov, \emph{{Quantum Loop Corrections in the Modified
  Gravity Model of Starobinsky Inflation with Primordial Black Hole
  Production}}, \href{https://doi.org/10.3390/universe10090354}{\emph{Universe}
  {\bfseries 10} (2024) 354},
  [\href{https://arxiv.org/abs/2402.02934}{{\ttfamily 2402.02934}}].

\bibitem{Ballesteros:2024zdp}
G.~Ballesteros and J.~G. Egea, \emph{{One-loop power spectrum in ultra
  slow-roll inflation and implications for primordial black hole dark matter}},
  \href{https://doi.org/10.1088/1475-7516/2024/07/052}{\emph{JCAP} {\bfseries
  07} (2024) 052}, [\href{https://arxiv.org/abs/2404.07196}{{\ttfamily
  2404.07196}}].

\bibitem{Firouzjahi:2024sce}
H.~Firouzjahi, \emph{{Two-Loop Corrections in Power Spectrum in Models of
  Inflation with Primordial Black Hole Formation}},
  \href{https://doi.org/10.3390/universe10120456}{\emph{Universe} {\bfseries
  10} (2024) 456}, [\href{https://arxiv.org/abs/2411.10253}{{\ttfamily
  2411.10253}}].

\bibitem{Sheikhahmadi:2024peu}
H.~Sheikhahmadi and A.~Nassiri-Rad, \emph{{Renormalized one-Loop Corrections in
  Power Spectrum in USR Inflation}},
  \href{https://arxiv.org/abs/2411.18525}{{\ttfamily 2411.18525}}.

\bibitem{Frolovsky:2025qre}
D.~Frolovsky and S.~V. Ketov, \emph{{One-loop corrections to the E-type
  $\alpha$-attractor models of inflation and primordial black hole
  production}},  \href{https://arxiv.org/abs/2502.00628}{{\ttfamily
  2502.00628}}.

\bibitem{Kristiano:2024vst}
J.~Kristiano and J.~Yokoyama, \emph{{Comparing sharp and smooth transitions of
  the second slow-roll parameter in single-field inflation}},
  \href{https://doi.org/10.1088/1475-7516/2024/10/036}{\emph{JCAP} {\bfseries
  10} (2024) 036}, [\href{https://arxiv.org/abs/2405.12145}{{\ttfamily
  2405.12145}}].

\bibitem{Kristiano:2024ngc}
J.~Kristiano and J.~Yokoyama, \emph{{Generating large primordial fluctuations
  in single-field inflation for PBH formation}},
  \href{https://arxiv.org/abs/2405.12149}{{\ttfamily 2405.12149}}.

\bibitem{Inomata:2024lud}
K.~Inomata, \emph{{Superhorizon Curvature Perturbations Are Protected against
  One-Loop Corrections}},
  \href{https://doi.org/10.1103/PhysRevLett.133.141001}{\emph{Phys. Rev. Lett.}
  {\bfseries 133} (2024) 141001},
  [\href{https://arxiv.org/abs/2403.04682}{{\ttfamily 2403.04682}}].

\bibitem{Kawaguchi:2024rsv}
R.~Kawaguchi, S.~Tsujikawa and Y.~Yamada, \emph{{Proving the absence of large
  one-loop corrections to the power spectrum of curvature perturbations in
  transient ultra-slow-roll inflation within the path-integral approach}},
  \href{https://doi.org/10.1007/JHEP12(2024)095}{\emph{JHEP} {\bfseries 12}
  (2024) 095}, [\href{https://arxiv.org/abs/2407.19742}{{\ttfamily
  2407.19742}}].

\bibitem{Fumagalli:2024jzz}
J.~Fumagalli, \emph{{Absence of one-loop effects on large scales from small
  scales in non-slow-roll dynamics II: Quartic interactions and consistency
  relations}},  \href{https://arxiv.org/abs/2408.08296}{{\ttfamily
  2408.08296}}.

\bibitem{Firouzjahi:2025gja}
H.~Firouzjahi and B.~Nikbakht, \emph{{Non-Perturbative Hamiltonian and Higher
  Loop Corrections in USR Inflation}},
  \href{https://arxiv.org/abs/2502.09481}{{\ttfamily 2502.09481}}.

\bibitem{Kinney:2005vj}
W.~H. Kinney, \emph{{Horizon crossing and inflation with large eta}},
  \href{https://doi.org/10.1103/PhysRevD.72.023515}{\emph{Phys. Rev. D}
  {\bfseries 72} (2005) 023515},
  [\href{https://arxiv.org/abs/gr-qc/0503017}{{\ttfamily gr-qc/0503017}}].

\bibitem{Namjoo:2012aa}
M.~H. Namjoo, H.~Firouzjahi and M.~Sasaki, \emph{{Violation of non-Gaussianity
  consistency relation in a single field inflationary model}},
  \href{https://doi.org/10.1209/0295-5075/101/39001}{\emph{EPL} {\bfseries 101}
  (2013) 39001}, [\href{https://arxiv.org/abs/1210.3692}{{\ttfamily
  1210.3692}}].

\bibitem{Martin:2012pe}
J.~Martin, H.~Motohashi and T.~Suyama, \emph{{Ultra Slow-Roll Inflation and the
  non-Gaussianity Consistency Relation}},
  \href{https://doi.org/10.1103/PhysRevD.87.023514}{\emph{Phys. Rev. D}
  {\bfseries 87} (2013) 023514},
  [\href{https://arxiv.org/abs/1211.0083}{{\ttfamily 1211.0083}}].

\bibitem{Chen:2013aj}
X.~Chen, H.~Firouzjahi, M.~H. Namjoo and M.~Sasaki, \emph{{A Single Field
  Inflation Model with Large Local Non-Gaussianity}},
  \href{https://doi.org/10.1209/0295-5075/102/59001}{\emph{EPL} {\bfseries 102}
  (2013) 59001}, [\href{https://arxiv.org/abs/1301.5699}{{\ttfamily
  1301.5699}}].

\bibitem{Morse:2018kda}
M.~J.~P. Morse and W.~H. Kinney, \emph{{Large-$\eta$ constant-roll inflation is
  never an attractor}},
  \href{https://doi.org/10.1103/PhysRevD.97.123519}{\emph{Phys. Rev. D}
  {\bfseries 97} (2018) 123519},
  [\href{https://arxiv.org/abs/1804.01927}{{\ttfamily 1804.01927}}].

\bibitem{Lin:2019fcz}
W.-C. Lin, M.~J.~P. Morse and W.~H. Kinney, \emph{{Dynamical Analysis of
  Attractor Behavior in Constant Roll Inflation}},
  \href{https://doi.org/10.1088/1475-7516/2019/09/063}{\emph{JCAP} {\bfseries
  09} (2019) 063}, [\href{https://arxiv.org/abs/1904.06289}{{\ttfamily
  1904.06289}}].

\bibitem{Dimopoulos:2017ged}
K.~Dimopoulos, \emph{{Ultra slow-roll inflation demystified}},
  \href{https://doi.org/10.1016/j.physletb.2017.10.066}{\emph{Phys. Lett. B}
  {\bfseries 775} (2017) 262--265},
  [\href{https://arxiv.org/abs/1707.05644}{{\ttfamily 1707.05644}}].

\bibitem{Chen:2013eea}
X.~Chen, H.~Firouzjahi, E.~Komatsu, M.~H. Namjoo and M.~Sasaki, \emph{{In-in
  and $\delta N$ calculations of the bispectrum from non-attractor single-field
  inflation}}, \href{https://doi.org/10.1088/1475-7516/2013/12/039}{\emph{JCAP}
  {\bfseries 12} (2013) 039},
  [\href{https://arxiv.org/abs/1308.5341}{{\ttfamily 1308.5341}}].

\bibitem{Akhshik:2015nfa}
M.~Akhshik, H.~Firouzjahi and S.~Jazayeri, \emph{{Effective Field Theory of
  non-Attractor Inflation}},
  \href{https://doi.org/10.1088/1475-7516/2015/07/048}{\emph{JCAP} {\bfseries
  07} (2015) 048}, [\href{https://arxiv.org/abs/1501.01099}{{\ttfamily
  1501.01099}}].

\bibitem{Akhshik:2015rwa}
M.~Akhshik, H.~Firouzjahi and S.~Jazayeri, \emph{{Cosmological Perturbations
  and the Weinberg Theorem}},
  \href{https://doi.org/10.1088/1475-7516/2015/12/027}{\emph{JCAP} {\bfseries
  12} (2015) 027}, [\href{https://arxiv.org/abs/1508.03293}{{\ttfamily
  1508.03293}}].

\bibitem{Mooij:2015yka}
S.~Mooij and G.~A. Palma, \emph{{Consistently violating the non-Gaussian
  consistency relation}},
  \href{https://doi.org/10.1088/1475-7516/2015/11/025}{\emph{JCAP} {\bfseries
  11} (2015) 025}, [\href{https://arxiv.org/abs/1502.03458}{{\ttfamily
  1502.03458}}].

\bibitem{Bravo:2017wyw}
R.~Bravo, S.~Mooij, G.~A. Palma and B.~Pradenas, \emph{{A generalized
  non-Gaussian consistency relation for single field inflation}},
  \href{https://doi.org/10.1088/1475-7516/2018/05/024}{\emph{JCAP} {\bfseries
  05} (2018) 024}, [\href{https://arxiv.org/abs/1711.02680}{{\ttfamily
  1711.02680}}].

\bibitem{Finelli:2017fml}
B.~Finelli, G.~Goon, E.~Pajer and L.~Santoni, \emph{{Soft Theorems For
  Shift-Symmetric Cosmologies}},
  \href{https://doi.org/10.1103/PhysRevD.97.063531}{\emph{Phys. Rev. D}
  {\bfseries 97} (2018) 063531},
  [\href{https://arxiv.org/abs/1711.03737}{{\ttfamily 1711.03737}}].

\bibitem{Passaglia:2018ixg}
S.~Passaglia, W.~Hu and H.~Motohashi, \emph{{Primordial black holes and local
  non-Gaussianity in canonical inflation}},
  \href{https://doi.org/10.1103/PhysRevD.99.043536}{\emph{Phys. Rev. D}
  {\bfseries 99} (2019) 043536},
  [\href{https://arxiv.org/abs/1812.08243}{{\ttfamily 1812.08243}}].

\bibitem{Pi:2022ysn}
S.~Pi and M.~Sasaki, \emph{{Logarithmic Duality of the Curvature
  Perturbation}},
  \href{https://doi.org/10.1103/PhysRevLett.131.011002}{\emph{Phys. Rev. Lett.}
  {\bfseries 131} (2023) 011002},
  [\href{https://arxiv.org/abs/2211.13932}{{\ttfamily 2211.13932}}].

\bibitem{Ozsoy:2021pws}
O.~\"Ozsoy and G.~Tasinato, \emph{{Consistency conditions and primordial black
  holes in single field inflation}},
  \href{https://doi.org/10.1103/PhysRevD.105.023524}{\emph{Phys. Rev. D}
  {\bfseries 105} (2022) 023524},
  [\href{https://arxiv.org/abs/2111.02432}{{\ttfamily 2111.02432}}].

\bibitem{Firouzjahi:2023xke}
H.~Firouzjahi and A.~Riotto, \emph{{Sign of non-Gaussianity and the primordial
  black holes abundance}},
  \href{https://doi.org/10.1103/PhysRevD.108.123504}{\emph{Phys. Rev. D}
  {\bfseries 108} (2023) 123504},
  [\href{https://arxiv.org/abs/2309.10536}{{\ttfamily 2309.10536}}].

\bibitem{Namjoo:2023rhq}
M.~H. Namjoo, \emph{{One consistency relation for all single-field inflationary
  models}}, \href{https://doi.org/10.1088/1475-7516/2024/05/041}{\emph{JCAP}
  {\bfseries 05} (2024) 041},
  [\href{https://arxiv.org/abs/2311.12777}{{\ttfamily 2311.12777}}].

\bibitem{Namjoo:2024ufv}
M.~H. Namjoo and B.~Nikbakht, \emph{{Non-Gaussianity consistency relations and
  their consequences for the peaks}},
  \href{https://doi.org/10.1088/1475-7516/2024/08/005}{\emph{JCAP} {\bfseries
  08} (2024) 005}, [\href{https://arxiv.org/abs/2401.12958}{{\ttfamily
  2401.12958}}].

\bibitem{Cai:2018dkf}
Y.-F. Cai, X.~Chen, M.~H. Namjoo, M.~Sasaki, D.-G. Wang and Z.~Wang,
  \emph{{Revisiting non-Gaussianity from non-attractor inflation models}},
  \href{https://doi.org/10.1088/1475-7516/2018/05/012}{\emph{JCAP} {\bfseries
  05} (2018) 012}, [\href{https://arxiv.org/abs/1712.09998}{{\ttfamily
  1712.09998}}].

\bibitem{Bassett:2005xm}
B.~A. Bassett, S.~Tsujikawa and D.~Wands, \emph{{Inflation dynamics and
  reheating}}, \href{https://doi.org/10.1103/RevModPhys.78.537}{\emph{Rev. Mod.
  Phys.} {\bfseries 78} (2006) 537--589},
  [\href{https://arxiv.org/abs/astro-ph/0507632}{{\ttfamily
  astro-ph/0507632}}].

\bibitem{Abolhasani:2019cqw}
A.~A. Abolhasani, H.~Firouzjahi, A.~Naruko and M.~Sasaki, \emph{{Delta N
  Formalism in Cosmological Perturbation Theory}}.
\newblock WSP, 2, 2019, \href{https://doi.org/10.1142/10953}{10.1142/10953}.

\bibitem{Ivanov:1994pa}
P.~Ivanov, P.~Naselsky and I.~Novikov, \emph{{Inflation and primordial black
  holes as dark matter}},
  \href{https://doi.org/10.1103/PhysRevD.50.7173}{\emph{Phys. Rev. D}
  {\bfseries 50} (1994) 7173--7178}.

\bibitem{Garcia-Bellido:2017mdw}
J.~Garcia-Bellido and E.~Ruiz~Morales, \emph{{Primordial black holes from
  single field models of inflation}},
  \href{https://doi.org/10.1016/j.dark.2017.09.007}{\emph{Phys. Dark Univ.}
  {\bfseries 18} (2017) 47--54},
  [\href{https://arxiv.org/abs/1702.03901}{{\ttfamily 1702.03901}}].

\bibitem{Germani:2017bcs}
C.~Germani and T.~Prokopec, \emph{{On primordial black holes from an inflection
  point}}, \href{https://doi.org/10.1016/j.dark.2017.09.001}{\emph{Phys. Dark
  Univ.} {\bfseries 18} (2017) 6--10},
  [\href{https://arxiv.org/abs/1706.04226}{{\ttfamily 1706.04226}}].

\bibitem{Biagetti:2018pjj}
M.~Biagetti, G.~Franciolini, A.~Kehagias and A.~Riotto, \emph{{Primordial Black
  Holes from Inflation and Quantum Diffusion}},
  \href{https://doi.org/10.1088/1475-7516/2018/07/032}{\emph{JCAP} {\bfseries
  07} (2018) 032}, [\href{https://arxiv.org/abs/1804.07124}{{\ttfamily
  1804.07124}}].

\bibitem{Khlopov:2008qy}
M.~Y. Khlopov, \emph{{Primordial Black Holes}},
  \href{https://doi.org/10.1088/1674-4527/10/6/001}{\emph{Res. Astron.
  Astrophys.} {\bfseries 10} (2010) 495--528},
  [\href{https://arxiv.org/abs/0801.0116}{{\ttfamily 0801.0116}}].

\bibitem{Ozsoy:2023ryl}
O.~\"Ozsoy and G.~Tasinato, \emph{{Inflation and Primordial Black Holes}},
  \href{https://doi.org/10.3390/universe9050203}{\emph{Universe} {\bfseries 9}
  (2023) 203}, [\href{https://arxiv.org/abs/2301.03600}{{\ttfamily
  2301.03600}}].

\bibitem{Byrnes:2021jka}
C.~T. Byrnes and P.~S. Cole, \emph{{Lecture notes on inflation and primordial
  black holes}},  12, 2021, \href{https://arxiv.org/abs/2112.05716}{{\ttfamily
  2112.05716}}.

\bibitem{Escriva:2022duf}
A.~Escriv\`a, F.~Kuhnel and Y.~Tada, \emph{{Primordial Black Holes}},
  \href{https://arxiv.org/abs/2211.05767}{{\ttfamily 2211.05767}}.

\bibitem{Pi:2024jwt}
S.~Pi, \emph{{Non-Gaussianities in primordial black hole formation and induced
  gravitational waves}},  \href{https://arxiv.org/abs/2404.06151}{{\ttfamily
  2404.06151}}.

\bibitem{Green:2024hbw}
D.~Green, K.~Gupta and Y.~Huang, \emph{{A Goldstone boson equivalence for
  inflation}}, \href{https://doi.org/10.1007/JHEP09(2024)117}{\emph{JHEP}
  {\bfseries 09} (2024) 117},
  [\href{https://arxiv.org/abs/2403.05274}{{\ttfamily 2403.05274}}].

\bibitem{Weinberg:1995mt}
S.~Weinberg, \emph{{The Quantum theory of fields. Vol. 1: Foundations}}.
\newblock Cambridge University Press, 6, 2005,
  \href{https://doi.org/10.1017/CBO9781139644167}{10.1017/CBO9781139644167}.

\bibitem{Weinberg:2005vy}
S.~Weinberg, \emph{{Quantum contributions to cosmological correlations}},
  \href{https://doi.org/10.1103/PhysRevD.72.043514}{\emph{Phys. Rev. D}
  {\bfseries 72} (2005) 043514},
  [\href{https://arxiv.org/abs/hep-th/0506236}{{\ttfamily hep-th/0506236}}].

\bibitem{Kapusta:2007xjq}
J.~I. Kapusta and C.~Gale, \emph{{Finite-Temperature Field Theory : Principles
  and Applications, 2nd edition}}.
\newblock Cambridge University Press, 2007,
  \href{https://doi.org/10.1017/9781009401968}{10.1017/9781009401968}.

\bibitem{Leblond:2010yq}
L.~Leblond and E.~Pajer, \emph{{Resonant Trispectrum and a Dozen More
  Primordial N-point functions}},
  \href{https://doi.org/10.1088/1475-7516/2011/01/035}{\emph{JCAP} {\bfseries
  01} (2011) 035}, [\href{https://arxiv.org/abs/1010.4565}{{\ttfamily
  1010.4565}}].

\bibitem{Behbahani:2011it}
S.~R. Behbahani, A.~Dymarsky, M.~Mirbabayi and L.~Senatore, \emph{{(Small)
  Resonant non-Gaussianities: Signatures of a Discrete Shift Symmetry in the
  Effective Field Theory of Inflation}},
  \href{https://doi.org/10.1088/1475-7516/2012/12/036}{\emph{JCAP} {\bfseries
  12} (2012) 036}, [\href{https://arxiv.org/abs/1111.3373}{{\ttfamily
  1111.3373}}].

\bibitem{Creminelli:2024cge}
P.~Creminelli, S.~Renaux-Petel, G.~Tambalo and V.~Yingcharoenrat,
  \emph{{Non-perturbative wavefunction of the universe in inflation with
  (resonant) features}},
  \href{https://doi.org/10.1007/JHEP03(2024)010}{\emph{JHEP} {\bfseries 03}
  (2024) 010}, [\href{https://arxiv.org/abs/2401.10212}{{\ttfamily
  2401.10212}}].

\end{thebibliography}\endgroup

\bibliographystyle{JHEP}

\end{document}